\pdfoutput=1
\documentclass[twocolumns,traditabstract]{aa}
\usepackage{graphicx}
\usepackage{amsmath}
\usepackage{amssymb}
\usepackage{latexsym}
\usepackage{graphicx}
\usepackage{gensymb}
\usepackage{xcolor}
\usepackage{enumerate}

\usepackage{txfonts}
\def\ltsima{$\; \buildrel < \over \sim \;$}
\def\simlt{\lower.5ex\hbox{\ltsima}}
\def\gtsima{$\; \buildrel > \over \sim \;$}
\def\simgt{\lower.5ex\hbox{\gtsima}}

\begin{document}
   \title{The multi-phase winds of Markarian 231: from the hot, nuclear, ultra-fast wind  to the galaxy-scale, molecular outflow
   \thanks{
Based on observations carried out with the IRAM Plateau de Bure Interferometer. IRAM is supported by INSU/CNRS (France), MPG (Germany) and IGN (Spain), 
and with {\it Chandra} and {\it NuSTAR} observatories.
}
}

   \author{C. Feruglio \inst{1,2,3}
          \and
	F. Fiore  \inst{3} 
        \and
       S. Carniani\inst{4,5,6}
	\and
            E. Piconcelli \inst{3}
           \and 
	L. Zappacosta \inst{3} 
	\and
           A. Bongiorno \inst{3}
           \and 
           C. Cicone \inst{5,6}
	\and 
	R. Maiolino \inst{5,6}
	\and 
	A. Marconi \inst{4}
	\and 
	N. Menci \inst{3}
	\and 
	S. Puccetti \inst{7,3}
	\and 
	S. Veilleux \inst{8,9} 
           }

   \institute{
   Scuola Normale Superiore, Piazza dei Cavalieri 7, 56126 Pisa, Italy, 
    \email{chiara.feruglio@sns.it}
   \and 
   IRAM -- Institut de RadioAstronomie Millim\'etrique, 
 300 rue de la Piscine, 38406 Saint Martin d'H\'eres, France
         \and
            INAF -- Osservatorio Astronomico di Roma, via Frascati 33, 00040 Monteporzio Catone, Italy
            \and
          Dipartimento di Fisica e Astronomia, Universit\'a di Firenze, Via G. Sansone 1, I-50019, Sesto Fiorentino (Firenze), Italy 
	\and
            Cavendish Laboratory, University of Cambridge, 19 J. J. Thomson Ave., Cambridge CB3 0HE, UK
          \and
	Kavli Insitute for Cosmology, University of Cambridge, Madingley Road, Cambridge CB3 0HA, UK
          \and 
           ASDC-ASI, Via del Politecnico, I-00133 Roma, Italy
          \and
           Department of Astronomy and CRESST, University of Maryland, College Park, MD, 20742, USA
           \and
           Joint Space Science Institute, University of Maryland, College Park, MD20742, USA
             }

%\date{.....}

% \abstract{}{}{}{}{} 
% 5 {} token are mandatory

 \abstract
{Mrk 231 is a nearby  ultra-luminous IR galaxy exhibiting a kpc-scale, multi-phase AGN-driven outflow. This galaxy represents the best target to investigate in detail the morphology and energetics of powerful outflows, as well as their still poorly-understood expansion mechanism and impact on the host galaxy.
In this work,  we present the best sensitivity and angular resolution maps of the molecular disk and outflow of Mrk~231, as traced by CO(2-1) and (3-2) observations obtained with the IRAM/PdBI. In addition, we analyze archival deep {\it Chandra} and {\it NuSTAR} X-ray observations.
We use this unprecedented combination of multi-wavelength datasets to constrain the physical properties of both the molecular disk and outflow, the presence of a highly-ionized ultra-fast nuclear wind, and their connection.
%The tilted inner ($\sim$100 pc) molecular disk, discovered by using dense gas tracers, is also seen in our high angular resolution CO(2-1) maps.
The molecular CO(2-1) outflow has a size of $\sim1$ kpc, and extends in all directions around the nucleus, being more prominent along the south-west to north-east direction, suggesting a wide-angle biconical geometry. 
The maximum projected velocity of the outflow is nearly constant out to $\sim1$ kpc, thus implying that the density of the outflowing material must decrease from the nucleus outwards as $\sim r^{-2}$. 
This suggests that either a large part of the gas leaves the flow during its expansion or that the bulk of the outflow has not yet reached out to $\sim 1$ kpc, thus implying a limit on its age of $\sim 1$ Myr. 
Mapping the mass and energy rates of the molecular outflow yields $\rm \dot M_{OF}=[ 500-1000]~ M_{\odot}~yr^{-1}$ and $\rm \dot E_{kin,OF}=[7-10]\times 10^{43}$ erg s$^{-1}$.
The total kinetic energy of the outflow is $\rm E_{kin,OF}$ is of the same order of  the total energy of the molecular disk, $\rm E_{disk}$.
Remarkably, our analysis of the  X-ray data reveals a nuclear ultra-fast outflow (UFO) with velocity $-20000$ km s$^{-1}$, $\rm \dot M_{UFO}=[0.3- 2.1] ~M_\odot yr^{-1}$, and momentum load $\rm \dot P_{UFO}/\dot P_{rad}=[0.2-1.6]$.
We find $\rm \dot E_{kin,UFO}\sim \dot E_{kin,OF}$ as predicted for outflows undergoing an energy conserving expansion.
This suggests that most of the UFO kinetic energy is transferred to mechanical energy of the kpc-scale outflow, strongly supporting that the 
energy released during accretion of matter onto super-massive black holes is the ultimate driver of giant massive outflows.
The momentum flux $\rm \dot P_{OF}$ derived for the large scale outflows in Mrk 231 enables us to estimate a momentum boost  $\rm \dot P_{OF}/\dot P_{UFO}\approx [30-60]$.
The ratios $\rm \dot E_{kin, UFO}/L_{bol,AGN} =[ 1-5]\%$ and $\rm \dot E_{kin,OF}/L_{bol,AGN} = [1-3]\%$ agree with the requirements  of the most popular models of AGN feedback.
}

%assuming constant velocity and constant emissivity, the fact that the receiding emission is brighter than the approaching emission, the cone is mainly oriented towards the


\titlerunning{The multi-phase winds of Mrk 231}
\authorrunning{C. Feruglio et al.}
 \maketitle

\section{Introduction}

Luminous AGN are likely to play a key role in the evolution of their host galaxies through powerful winds and outflows. 
In particular, during the earliest, dust obscured phase of AGN activity, characterized by a large amount of molecular gas 
available to fuel black hole accretion, the huge nuclear energy output enables the production of AGN-driven outflows.
These can perturb and possibly expel most of the gas out of the host galaxy, therefore leading to the quenching of star-formation. 
Current theoretical models of galaxy evolution include AGN feedback to account for galaxy colours and the lack of a large population of local massive star forming galaxies
(SFG, Granato et al. 2001, 2004, Di Matteo et al. 2005, Menci et al. 2006, 2008, Bower et al. 2006, Croton et al. 2006). 

Observations across the electromagnetic spectrum reveal that AGN winds are widespread, suggesting that they cover large solid angles and an extremely broad range of gas ionization parameter (Elvis 2000 and references therein). In X-rays mildly ionized warm absorbers with velocities of $\sim$500-1000 km s$^{-1}$ are observed in more than half  of AGN (Piconcelli et al. 2005; Blustin et al. 2005; McKernan et al. 2007). Furthermore, there is growing evidence for the existence of ultra-fast outflows (UFOs) via the detection of highly ionized Fe K-shell absorption arising from a gas moving at velocities up to 0.2--0.4$c$ 
%and   ionization parameters $\xi={L_{ion} \over n r^2}\sim 10^3-10^4$
 (Pounds et al. 2003; Reeves et al. 2009; Tombesi et al. 2011, 2015 and references therein).
Optical/UV Broad Absorption Line  (BAL) systems associated to ionized outflowing ($-2000 \div -20000~ \rm km~s^{-1}$) gas has been observed in 20-40\% of QSOs (Dai et al. 2008), with some remarkable examples reaching up to hundred of pc away from the nucleus (e.g., Borguet et al. 2013).
Broad [OIII] emission lines with velocities up to a few thousands km s$^{-1}$ are commonly found in AGN (Zhang et al. 2011; Shen \& Ho, 2014), 
and ionized and neutral gas with similar velocities is found on galactic scale (1-30 kpc) in a few tens AGN from the local Universe up to z=2-3 (Cano-Diaz et al. 2012, Rupke \& Veilleux 2011, 2013; Harrison et al. 2014; Nesvadba et al. 2008; Harrison et al. 2012; Genzel et al. 2014, Cresci et al. 2015). 
Radio observations have also detected many neutral atomic gas outflows with velocities up to 1500 km s$^{-1}$  (Morganti et al. 2005; Teng et al. 2013). Finally, fast (up to 1000-2000 km s$^{-1}$) molecular gas, extending on kpc scales is found in a few dozen, mostly highly obscured, AGN in dusty SFGs (Feruglio et al. 2010,  Cicone et al. 2014, and Veilleux et al. 2013 for compilations of molecular outflows). 

Two main scenarios for quasar-mode winds and feedback have been developed so far. 
The first is a two-stage mechanism (Lapi et al. 2005; Menci et al. 2008; Zubovas \& King 2012; Faucher-Giguere \& Quataert 2012), 
where a (semi)relativistic wind (such as the UFOs detected in the X-ray band), accelerated by the AGN radiation, shocks against the galaxy ISM. 
%These UFOs may be tentatively identified with the (semi)relativistic wind of Zubovas \& King (2012).
The shocked gas can reach temperatures of $10^7$ K or larger. Hot gas extended on kpc scales has been detected in galaxies with galactic-scale outflows (Feruglio et al. 2013, Wang et al. 2014, Veilleux et al. 2014).  
If the cooling time of the post shock gas is smaller than the dynamical time, $\rm t_{cool}< t_{dyn}$, then the gas can cool to $\sim 10^{4}$ K, emitting typical warm ionized gas lines ([OIII], H$\alpha$, etc). Further cooling leads to the formation of $\rm H_2$, which can further enhance the cooling rate, promptly leading to the formation of more complex molecules, and a multi-phase, galaxy-scale outflow with a momentum boost $\rm \dot P_{OF}> 10\times \dot P_{rad}$ should develop. 
Simulations of AGN feedback triggered by cold precipitation produce indeed a multi-phase structure of the gas in the galaxy cores (Gaspari et al. 2013).

In the second scenario, outflows are radiatively accelerated by  much slower, 'gentle'  winds, in which the momentum boost is at most $\rm \dot P_{OF}\lesssim 5\times \dot P_{rad}$ (Thompson et al. 2014). 
This may prevent gas clouds from escaping the halo gravitational potential, hence making them rain back onto the galaxy. 
Self-shielding of the galactic disk dusty clouds may further minimize AGN feedback.

The geometry of quasar-driven outflows as well as their  energy and momentum transport mechanisms are still poorly constrained by observations. 
This prevents both a clean discrimination between the two scenarios, and the development of detailed multi phase feedback models. 
This work aims at filling this gap by means of high spatial resolution CO mapping and deep X-ray spectroscopy of Mrk 231. 
Mrk 231 has been known for long as the closest ULIRG/QSO, with a total infrared luminosity of $\rm 3.6 \times 10^{12}~L_{\odot}$ (Sanders et al. 2003), AGN bolometric luminosity of $5\times 10^{45}$ erg $\rm s^{-1}$, and a star formation rate of 140 $\rm M_\odot ~yr^{-1}$ (Veilleux et al. 2009).  
It is an iron low-ionization broad absorption-line quasars (FeLoBAL), and it displays nuclear neutral and ionized outflows in several optical and UV tracers (Lipari et al. 2009, Veilleux et al. 2014), and a powerful neutral outflow extended on at least 3 kpc (Rupke \& Veilleux 2011). 
This is tracing a wide-angle conical outflow proceeding at an uniform speed of $-1000$ km s$^{-1}$ along the minor axis of the molecular disk. 
The first prototypical massive molecular outflow, extended on kpc scale, was discovered in this galaxy (Feruglio et al. 2010, Fischer et al. 2010, Aalto et al. 2012, 2014, Gonzalez-Alfonso et al. 2014). 
These observations leave several questions open, i.e. (i) what are the size and the geometry of the molecular outflow, 
(ii) whether and how this outflow is associated with the atomic neutral and ionized ones,  (iii) which is the momentum boost of the molecular outflow, (iv) is there evidence of any fast highly ionized nuclear wind in Mrk~231?

This work is a study of the multi-phase winds in Mrk~231 (z=0.04217, $\rm D_L$=187.6 Mpc, scale 0.837 kpc/\arcsec).  
Specifically, Section 2 presents the (sub)millimeter observations of molecular gas traced by CO(2-1) and (3-2), obtained with the IRAM/PdBI. 
Section 3 presents an analysis of the kinematics of the nuclear region of Mrk 231, including the molecular disk and the outflow, and spatially resolved mass flow rates and outflow kinetic power.
Section 4 presents an analysis of archival X-ray data from {\it Chandra} and {\it NuSTAR}, that led us to detect a nuclear, highly ionized UFO. 
In Section 5 the relation between the spatially-extended molecular outflow and the UFO is discussed in the framework of models for energy-conserving large-scale winds driven by AGN.

%tabella observations
\begin{table*}
\begin{center}
\caption{Plateau de Bure Interferometer observations of Mrk 231 used in this work.}
\begin{scriptsize}
\label{tab1}
\begin{tabular}{cccccccc}
\hline
\multicolumn{1}{l} {Name}&
\multicolumn{1}{c} {Date}&
\multicolumn{1}{c} {Array}&
\multicolumn{1}{c} {Freq.}&
\multicolumn{1}{c} {Line}&
\multicolumn{1}{c} {On source time}&
\multicolumn{1}{c} {Clean beam}&
\multicolumn{1}{c} {Sensitivity}\\
  & &  & [GHz] & & [hr] & [arcsec] & [mJy/beam/20MHz]\\
 \hline 
DS1 & Feb 2013 & A & 221.21 & CO(2-1) & 5.2 & $0.48\times0.36$ & 1.3\\
DS2 & Sep-Nov 2010/Feb 2013 & ACD & 221.21 & CO(2-1) & 8.7 & $0.9\times0.8$ & 0.8 \\
 DS3   & Jan 2012 &   D       &  331.8       &  CO(3-2)     & 2.0   & $1.5\times1.3$  &  8.0\\        
          \hline
\end{tabular}
\end{scriptsize}
\end{center}
\end{table*}

\section{(Sub)-Millimeter Observations}
We use three data sets obtained with the IRAM Plateau de Bure Interferometer (PdBI). 
The first dataset (DS1) consists of observations carried out with the most extended (A) 6 antenna array configuration, which has a maximum baseline length of 760 m, at a frequency of 221.21 GHz, targeting the CO(2-1) line.  
Absolute flux calibration is based on the quasars 3C84 (13.2 Jy) and 1150+497 (2.6 Jy), which also served as amplitude/phase calibrator. 
The accuracy of the absolute flux scale is of the order of 20\%, while the relative flux calibration between the 4 available observations is better than 10\%. 
The on source time of DS1 after calibration is 5.2 hours. 
The synthesized beam is 0.48 by 0.36 arcsec, PA 73 deg ($\sim$400 pc at the redshift of the source z$=0.04217$).
The achieved visibility (thermal) noise is 1.3 mJy/beam in 20 MHz channels (corresponding to 27 km s$^{-1}$). 
%We used the 1 mm continuum emission to self-calibrate the data, in order to improve the dynamical range. 
Natural weighting and a simple cleaning algorithm (\textit{hogbom}) were used in order to minimize secondary lobes. 
The visibility tables and data cubes have been binned to a 20 MHz spectral resolution, which is our preferred trade-off between spectral sampling and sensitivity. 
%Clean maps were produced for each channel in order to optimize polygons. 

The second data set (DS2) has been obtained by merging the visibilities of DS1 with previous observations with short baselines reported in Cicone et al. (2012). The relative flux calibration of the short and long-baseline data is based on the quasar 3C84, which has varied from 7.7 Jy in 2010 to 13.2 Jy in 2013.  Based on the light curve of 3C84 around 220 GHz we can expect that the two data sets have consistent flux scales at the 20\% level. 
DS2 has an on source time of 8.7 hours, and $1\sigma$ sensitivity 0.8 mJy/beam in 20 MHz channels.  
Maps for DS2 were produced as detailed above, resulting in a clean beam of 0.9 by 0.8 arcsec, PA 57. 
%No tapering of the beam was applied because DS2 allows to compensate for the flux loss in the first dataset. 
 
The third data set (DS3) consists of an observation of CO(3-2) at 331.8 GHz done in January 2012, using the compact (D) array configuration. The maximum phase rms during the observation was about 50 deg,  water vapor around 1 mm, and system temperatures between 300 and 500 K. The quasar 1150+497 (0.96 Jy at 331.9 GHz) was used for phase/amplitude calibration. MWC349 was used for flux calibration and provides an accuracy of about 20\% at this frequency. 
The synthesized beam is 1.5 by 1.3 arcsec, with a PA of 84 deg. The 1$\sigma$ sensitivity is 8.0 mJy/beam per 20 MHz.

DS1 has the best angular resolution and will be used to locate and characterize the kinematics of the inner $\sim1 arcsec$ of the system. 
DS2 has the best sensitivity and will be used to map fainter and more diffuse emission.
%All three data sets, along with previous CO(1-0) data presented in Feruglio et al. (2010) and Cicone et al. (2012), are then joined together to derive the CO spectral line energy distribution (SLED). 
Table 1 summarizes the main parameters of the observations.

\begin{figure*}[t]
\centering
\includegraphics[width=\textwidth]{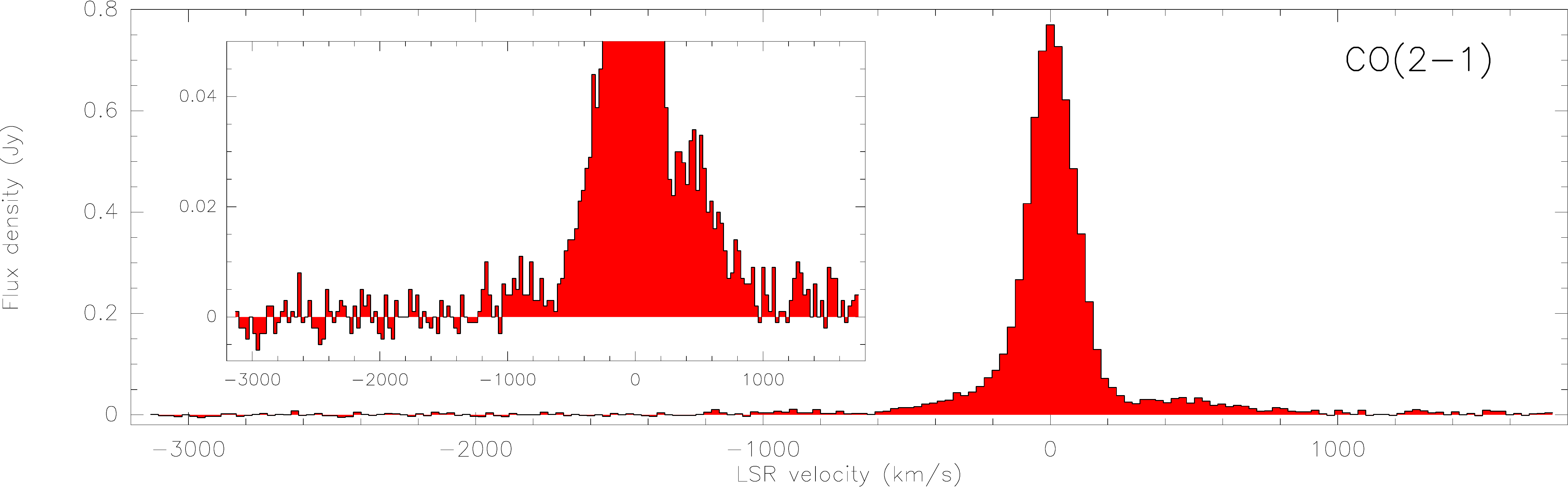}\\
\includegraphics[width=\textwidth]{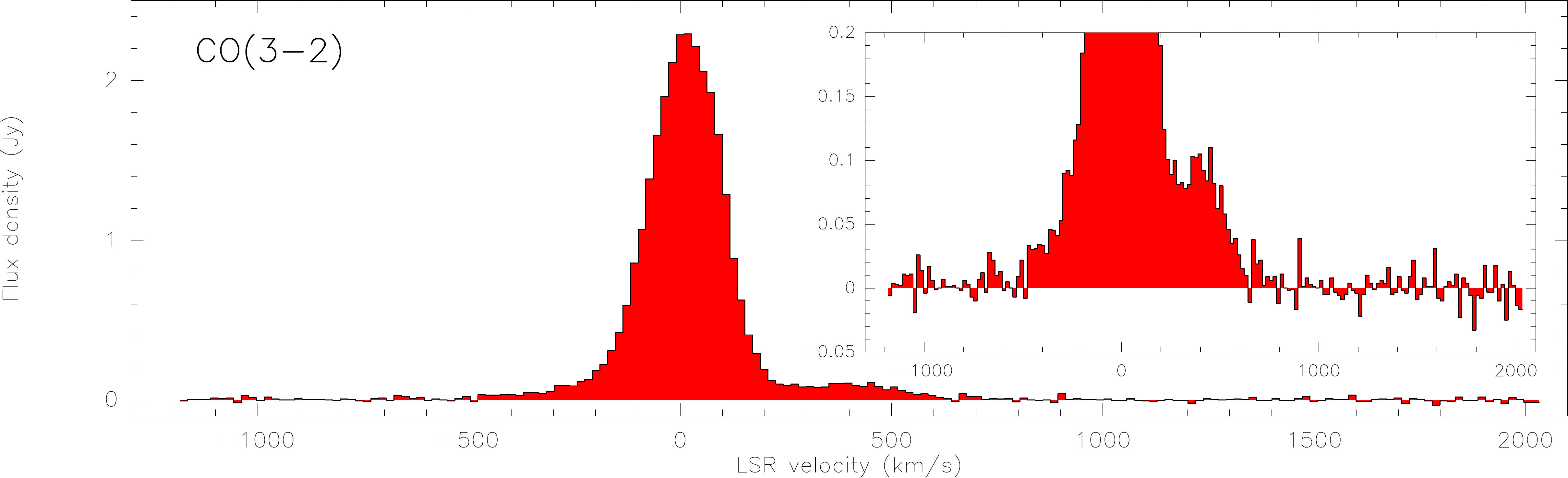}
\caption{ Continuum-subtracted spectra of CO(2-1) from DS1 (top panel) and of CO(3-2) (bottom panel). The plots display the full velocity range achieved by the observed bandwidth. Zoomed views  are shown in the insets. %extracted over a circular region of size 2 arcsec. 
}
\label{spectra}
\end{figure*}

\begin{figure}
\centering
\includegraphics[scale=0.28]{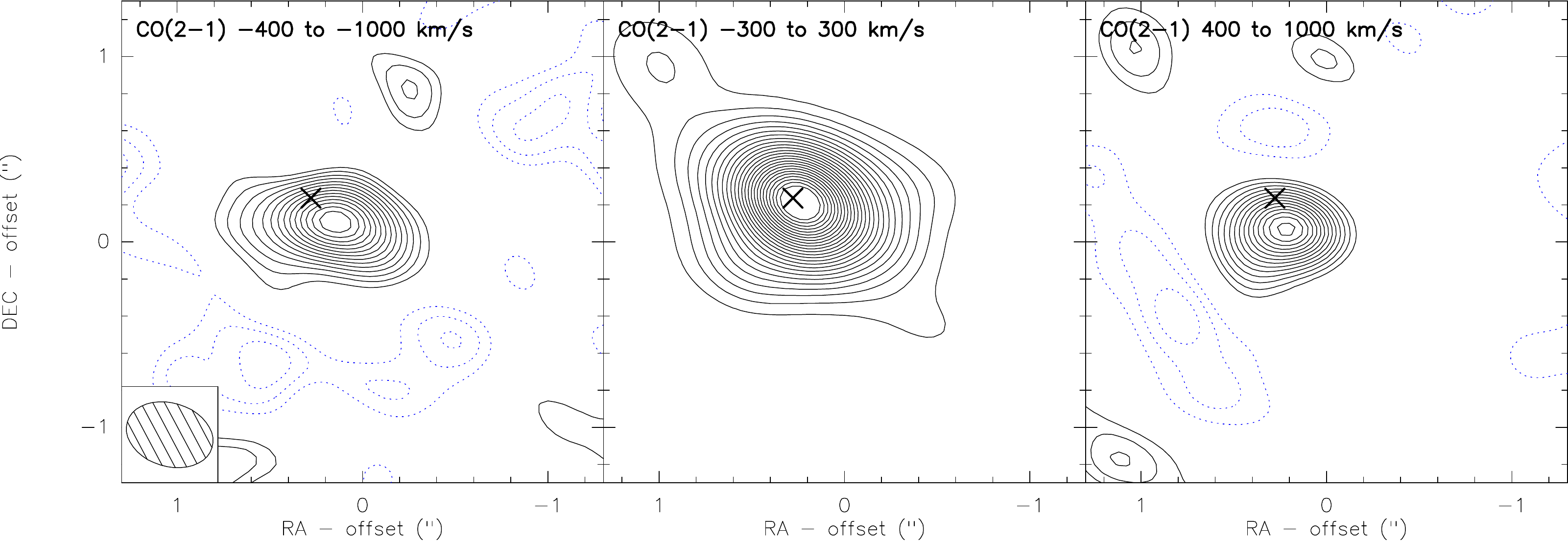}\\
\includegraphics[scale=0.28]{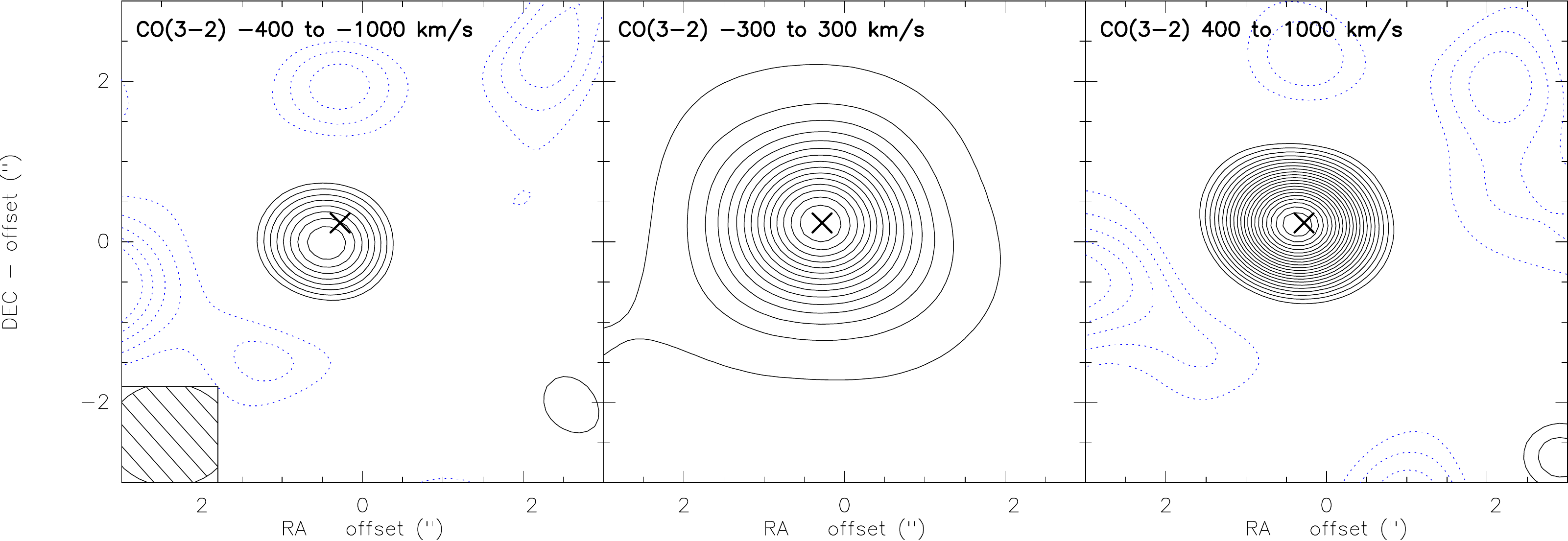}
\caption{Channel maps of CO(2-1) (DS1, upper panels) and CO(3-2) (lower panels) in three different velocity ranges (indicated by the labels).  The contours start from 4$\sigma$ and are in steps of 1$\sigma$ for the blue and red wing maps (right and left panels for both CO transitions, respectively). 
Negative contours are shown by dotted lines, and are in steps of 1$\sigma$, starting at $-1\sigma$.  
In the systemic component map (middle panels), contours are in steps of 10$\sigma$ starting from 10$\sigma$, for the sake of clarity.
The synthesized beams are shown by hatched areas for each data set. 
1$\sigma = 0.14~ \rm Jy~ km~ s^{-1}$ for CO(2-1), $= 0.72$ Jy km s$^{-1}$ for CO(3-2). 
The cross marks the position of the 1.4 GHz continuum as measured by the VLBI, $\alpha=$12:56:14.2339, $\delta=$56:52:25.237 (J2000, Ma et al. 1998). 
 }
 \label{channel-maps}
\end{figure}

\subsection{1.4 mm data}
The 1.4 mm continuum was estimated by averaging the visibilities in two spectral regions free of lines in the range $-3000$ to $-2000$ km s$^{-1}$, and 1500 to 1600 km s$^{-1}$. 
We tested different channel ranges to average the continuum visibilities, mainly on the blue side (1500 to 3000 km/s negative velocities).
The features seen at velocity $-600$ out to $-1000$ km/s, and those out to +1000 km/s, persist using different channel ranges to estimate the continuum.
The significance of spectral features beyond these limits are more difficult to assess. 
In particular, on the red side, where the spectral range is limited to +1800 km/s, the estimate of the continuum may be more uncertain, and features redder than +1000 km/s could be spurious.
A point source fit to the visibilities averaged as explained above, gives a 1.4 mm continuum flux density of $43.8 \pm 0.5$ mJy (Table \ref{tab2}), in agreement both with the recent interferometric measurement of Aalto et al. (2014), done with similar baselines, and with Downes \& Solomon (1998).

The continuum visibility table, derived as explained above, has been subtracted from the total uv data to produce the line table and the corresponding map. 
Figure 1, upper panel, shows the CO(2-1) spectrum, extracted in a circular region of diameter 1.5 arcsec centered on the emission peak on the map.
The peak flux of CO(2-1) is 0.73 Jy, from a Gaussian fit of the spectral line, and agrees within the flux accuracy ($\pm\%20$) with previous interferometric measurements of  Cicone et al. (2012), Downes \& Solomon (1998) (which had slightly larger beam), and with the flux measured with the IRAM 30 m dish (Solomon et al. 1997).
This suggests that little emission is filtered out by long baselines and that most of  the CO emission occurs within a 1.5 arcsec central region. 
The line profile is similar to that found in previous CO(1-0) and (2-1) observations (Feruglio et al. 2010, Cicone et al. 2012). 
CO(2-1) emission is detected out to about $\pm1000$ km s$^{-1}$.

%tabella continuo  
\begin{table}[b]
\caption{1.4 and 0.9 mm continua visibility fit parameters}
\begin{scriptsize}
\label{tab2}
\begin{center}
\begin{tabular}{lcccc}
\hline
\multicolumn{1}{c} {Wavelength}&
\multicolumn{1}{c} {Model}&
\multicolumn{1}{c} {RA} &
\multicolumn{1}{c} {DEC}&
\multicolumn{1}{c} {S$_{\nu, zero-spacing}$}\\
\hline
1.4 mm & point source&  12:56:14.23 & 56:52:25.24 & $43.8\pm0.5$ mJy\\
0.9 mm &point source& 12:56:14.24 & 56:52:25.23 & $72.0\pm2.0$ mJy\\ 
\hline
\end{tabular}
\end{center}
Note. - Errors: 
%for positions at 1.4 mm, $\pm$; at 0.9 mm, $\pm$; 
flux errors are statistical, additional $\pm20\%$ should be accounted for flux calibration. 
\end{scriptsize}
\end{table}

Figure 2 (upper panels) shows the channel contour maps of CO(2-1) emission integrated around the systemic velocity ($\pm 300$ km s$^{-1}$), in the blue (-400 to -1000 km s$^{-1}$), and red (400 to 1000 km s$^{-1}$) sides of the line. Both the systemic and the red/blue shifted emission are more extended than the synthesized beam and have a size up to $\sim1.5$ arcsec. 
We performed visibility fitting of these three CO(2-1) line components, systemic, blue and redshifted emission, after having averaged each visibility set in the corresponding spectral range (similarly to Cicone et al. (2012), and references therein).
The positions, fluxes and sizes so derived are reported in  Table \ref{tabvis} for each component, together with the respective spectral range and the adopted source model.
The visibility fits are independent of the synthesized beam and of the CLEAN algorithm, and there is no need to deconvolve the apparent size on the map. 
The integrated fluxes agree within the errors with the measurements of Cicone et al. (2012).  
The FWHM size of the systemic gas component is $0.86\pm0.01$ arcsec, corresponding to a physical scale of $\sim700$ pc. 
The bulk of the blue and red wing emission of CO(2-1) is compact, having FWHM sizes of about a beam by fitting a with Gaussian function. 
A fainter spatially extended structure is seen in the channel maps of the red  and blue wings (Fig. 3), which can hardly be fitted by a simple gaussian model, and, in the case of the receding component, is extended out to at least 1\arcsec ($\sim1$ kpc) north-east off the AGN. 
This component is also seen in HCN(1-0), on similar spatial scales and orientation (Aalto et al. 2012).
The positions of both the systemic CO emission and of the 1.4 mm continuum are consistent with that of the 1.4 GHz radio continuum measured by the VLBI, which traces the AGN emission.  %The offset between VLBI and PdBI positions  (0.05\arcsec) may be due to the incomplete uv coverage of PdBI data. 
The blue and the redshifted emissions are instead both offset from the AGN position in the south-western direction by  ($-0.2,-0.3$) and ($-0.2,-0.1$) arcsec, respectively. These offsets are about $\sim$10 larger than the position errors of our data,  i.e., $0.02\arcsec$ for the blue and $0.01\arcsec$ for the red component, respectively (Table \ref{tabvis}). 
The positions of the wing emission are consistent with those seen in HCN(1-0) (Aalto et al. 2012), and with those previously measured by Cicone et al. (2012)\footnote{Note that in Cicone et al. (2012) the phase reference center was  wrongly quoted as the VLBI position, which produces an apparent discrepancy in the location of the outflow between that and this work.}.

 \begin{figure*}[t]
\centering
\includegraphics[width=15cm]{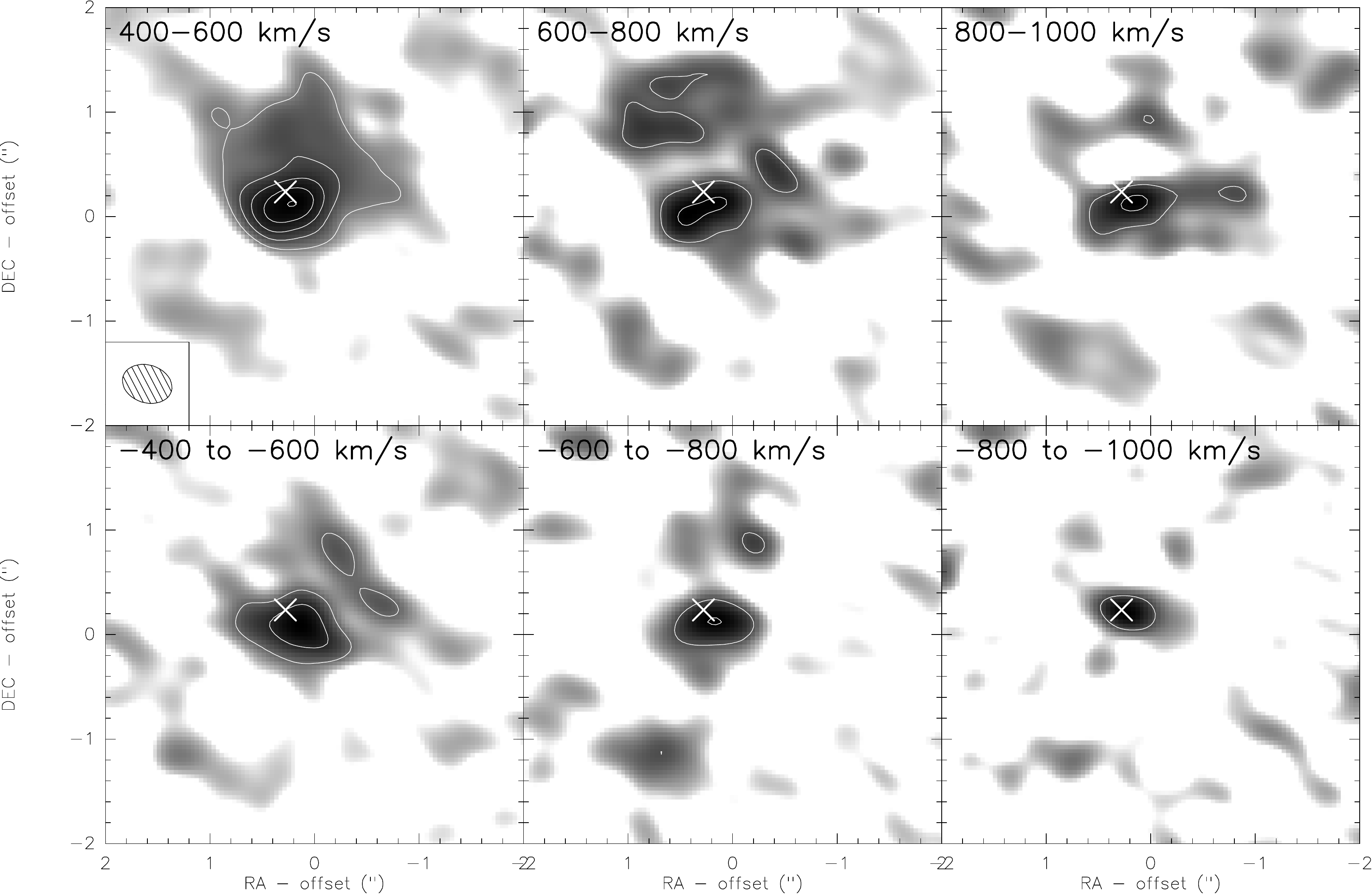}
\caption{Channel maps of CO(2-1) from DS1. Contours are in steps of 5$\sigma$, starting at 5$\sigma$. $1\sigma=0.09$ Jy km s$^{-1}$.
}
\label{channel-wings}
\end{figure*}

%\subsection{Blueshifted CO(2-1) emission}

We discuss in the following the possibility that the spectral feature seen between -1000 and -700 km s$^{-1}$ traces emission from a species different from CO. Specifically, this spectral feature peaks at a frequency of 221.864 ($-860$ km s$^{-1}$), which corresponds to that expected for $\rm ^{13}CS(5-4)$ (rest frame frequency 231.220 GHz). 
$^{13}$CS(5-4) was first detected in an extragalactic source in the 1 mm survey of Arp220 (Martin et al. 2011), and traces very dense gas. 
Its critical density is, in fact, of the order of $10^6$ cm$^{-3}$. 
The integrated emission is detected with an $8\sigma$ significance. Its FWHM is about 150 km s$^{-1}$, and the zero spacing flux is $2.5\pm0.3$ mJy, by fitting an unresolved source. 
$^{12}$CS(5-4) is undetected in the IRAM 30 m spectrum down to a 1$\sigma$ sensitivity of $\sim$1 mK (8.7 mJy, Zhang, PhD thesis). 
The isotopic ratio $\rm ^{12}C/^{13}C$ in Mrk 231 is about 40, based interferometric data of $^{12}$CO and $^{13}$CO J=1-0 (Feruglio et al. in preparation), which agrees with the lower limit given by Glenn \& Hunter (2001).  
Assuming that $\rm ^{13}CO$ comes from approximately the same region as $^{12}$CO(1-0), we can apply  this isotopic ratio to derive the expected strength of $^{13}$CS(5-4), which would be $<0.2$ mJy. 
This suggests that the feature can hardly be identified with $^{13}$CS(5-4), and we conclude, therefore,  that it is due to approaching CO(2-1).

\subsection{0.9 mm data}

The 0.9 mm continuum has been estimated by fitting in the uv plane visibilities averaged in spectral windows. 
We have tested the continuum estimate and subtraction by using three different spectral windows for visibility averaging. 
In particular, averaging the spectral windows $-1180$ to $-1000$ km/s, plus 1400 to 2000 km/s produces over-subtraction of the continuum red-wards of the line.
We find that the best overall continuum subtraction results from using the spectral window 1400 to 2000 km/s, which gives a continuum flux density of $72.0\pm 2.0$ mJy (Table \ref{tab2}). This is our adopted best solution. The continuum visibility table, produced as explained above, has been subtracted from the total visibilities in order to produce the line emission visibility table, which has been used to map the CO(3-2) emission (Figure 2, bottom panels).

The CO(3-2) spectrum, extracted from a circular aperture of 3 arcsec diameter around the map peak, is shown in Figure 1 (bottom panel). 
We measure a  flux density of  $658\pm5$ Jy km s$^{-1}$ for the CO(3-2) systemic component. 
The single dish flux from Wilson et al. (2008), $480\pm100$ Jy km s$^{-1}$, is slightly smaller than our estimate.  
%The interferometric flux, also from Wilson et al. (2008), is $308 \pm8$Jy km/ with a beam of $0.9\times1.0\arcsec$, inconsistent with our flux.  
A high velocity component is detected in CO(3-2) out to speeds of $\pm 1000$ km s$^{-1}$, similarly to CO(1-0) and (2-1). 
The CO(3-2)  profile is similar to that of CO(2-1), (1-0), showing an enhanced red wing compared to the blue one.  
We acknowledge that the systematic uncertainty on the continuum measurement can affect the significance of the faint features seen at large velocities (in particular blue-wards of the line at velocity $< -1000$ km/s), but features within $\pm800$ km/s from the line peak are persistent by adopting any of our tested spectral windows.
The positions of the systemic, the red and blue-shifted components of CO(3-2) are consistent with those derived for CO(2-1), within the uncertainty associated to the larger beam and to the coarser uv coverage of DS3 (Table \ref{tabvis}). 
%are offset compared to those of  CO(2-1).  This offset is certainly due to the larger error in the positions in CO(3-2) data. In fact, with a beam of $1.5\times1.3$ arcsec, the error in the position is 0.2\arcsec for the blue and 0.1\arcsec for the red wing, from the relation used above, but are likely larger due to the poor uv coverage of the CO(3-2) data (2 hours only spent on the source). 

In the following we discuss the possible contribution of  $\rm H^{13}CN(4-3)$ emission to the red wing of CO(3-2). 
Sakamoto et al. (2009) found a ratio CO(3-2) /$\rm H^{13}CN(4-3)$$ =$ 10 in NGC4418, which implies severe contamination of the CO(3-2) red wing for that source.  A ratio of $\sim10$ is excluded for Mrk 231 by the observed spectrum already (Fig. 1).  
Costagliola et al. (2011) measured for Mrk 231 a $\rm H^{13}CN(1-0) = 1.7$ Jy km s$^{-1}$ at a 3$\sigma$ level. %( $0.34\pm 0.1$ K km s$^{-1}$ ) 
From their spectrum, the line appears somewhat broader than the HCN(1-0), suggesting that  it can be blended with some other nearby lines. 
Indeed, emission from SO (86.1 GHz), and SO$_2$ (on the red side at 86.639 GHz) are expected within $\pm 300$ MHz .
These transitions are as bright as $\rm H^{13}CN$ in the Orion and SgrB2 star forming regions (Turner et al. 1989), which can in principle also be the case in Mrk 231.
Therefore the measure of Costagliola et al. (2011) must be considered as an upper limit. In fact, this would provide a ratio HCN/$\rm H^{13}CN(1-0)\sim 8$, 
based on the strength of HCN(1-0) measured by Aalto et al. (2012), i.e., a factor of five smaller than the isotopic $\rm ^{12}C/^{13}C$ ratio of 40 found by us. 
Adopting an isotopic ratio of 8 and $\rm HCN(4-3) = 65\pm13$ Jy km s$^{-1}$ (Papadopoulos et al. 2007),
we get $\rm H^{13}CN(4-3) = 8$ Jy km s$^{-1}$,  meaning that this would contribute 25\% of the integrated flux in the red wing of CO(3-2).
Adopting instead a ratio $\rm^{13}CO/^{12}CO=40$, the contribution to the CO(3-2) red wing would be less than 5\%.  

The estimates given above assume that $\rm H^{13}CN(4-3)$ is optically thick. Should this not be the case, its flux may be even smaller. 
Modeling with RADEX non-LTE radiative transport models (van der Tak et al. 2007) to reproduce flux of CO and HCN, suggests  instead that $\rm H^{13}CN(4-3)$ is optically thin.
In fact, assuming $\Delta \rm v = 60$ km s$^{-1}$ and typical temperatures and column densities ($\rm T_{kin} = 60$ K,  $\rm N_H=1\times 10^{16}$ cm$^{-2}$ for HCN and $1\times 10^{15}$ cm$^{-2}$ for $\rm H^{13}CN$), we get  $\tau \sim 1\times 10^{-2}$ for $\rm H^{13}CN$(4-3).   
In this case the flux of $\rm H^{13}CN$(4-3) would be very small and the contamination of the red wing of CO(3-2) negligible. 
The maps of the CO(3-2)  give an observed temperature of  $\rm T= 0.2$ K in the red wing. 
The HCN(4-3)  flux (65 Jy km s$^{-1}$, Papadoupulos et al. 2007), implies 0.325 Jy for a line width of 200 km s$^{-1}$. 
%If the beam is $1.5\arcsec ~\times 1.3\arcsec$ and 
Ignoring the frequency difference between 340 and 330 GHz, than we would get 5.7 K/Jy for the PdBI data, thus $\rm T(HCN(4-3)  =  1.85~ K$ (i.e. 0.325 Jy). 
Modeling with RADEX, and assuming  $\rm N(H^{13}CN) \sim 1\times 10^{15}$ cm$^{-2}$, $\rm n(H_2) \sim 1\times 10^{4}$ cm$^{-3}$, and $\rm T_{kin} = 60 ~K$, we get  $\rm T (H^{13}CN(4-3)) \sim  2.5\times 10^{2}$ K, and 
$\rm S (H^{13}CN(4-3) )= 4.4 ~mJy$, which in turn would give 0.88 Jy km s$^{-1}$ integrated intensity for a line width of 200 km s$^{-1}$.  
In conclusion, a strong contamination from $\rm H^{13}CN(4-3)$ is unlikely and most of the detected flux in the red wing of CO(3-2) is from high speed CO(3-2).

%tabella con uv fits  CO
\begin{table*}
\caption{Visibility fits of the CO(2-1) and CO(3-2)}
\begin{scriptsize}
\begin{center}
\begin{tabular}{llccllccc}
\hline
\multicolumn{1}{c} {Component}&
\multicolumn{1}{c} {Vel. range}&
\multicolumn{1}{c} {Dataset}&
\multicolumn{1}{c} {Model}&
\multicolumn{1}{c} {RA} &
\multicolumn{1}{c} {DEC}&
\multicolumn{1}{c} {S$_{\nu, zero-spacing}$}&
\multicolumn{1}{c} {$\int S_\nu dv$}&
\multicolumn{1}{c} {Source Size FWHM}\\
%\multicolumn{1}{c} {Luminosity}\\
& [km s$^{-1}$] & & & & & [mJy] & [Jy km s$^{-1}$] & [arcsec] \\
\hline
CO(2-1) Systemic & $-300\div$+300 & DS1 & expon. disk &  12:56:14.230 [$\pm0.003"$] & 56:52:25.200 [$\pm0.002$"] & $311.0\pm0.6$ & 186.6$\pm$3.8 & 0.86$\pm$0.01 \\ %& $3.9~10^9$\\ 
CO(2-1) Blue &  $-400\div -1000$ & DS1 &gauss & 12:56:14.22 [$\pm0.02"$] & 56:52:25.06 [$\pm0.02"$] & 6.9$\pm$0.1 & 4.2$\pm$0.5 & 0.35$\pm$0.07  \\%& $8.7~10^7$\\
CO(2-1) Red & $400\div 1000$ &  DS1 & gauss & 12:56:14.23 [$\pm0.01"$] & 56:52:25.04 [$\pm0.01"$] & 12.3$\pm$0.1 & 7.4$\pm$0.6  & 0.44$\pm$0.05 \\%&$1.5~10^8$\\
\hline
CO(3-2) Systemic &$-300\div$+300 & DS3 & expon. disk & 12:56:14.24[$\pm0.003"$] & 56:52:25.19[$\pm0.003"$] & $\rm 1096\pm9$ & $658\pm5.0$ & $1.150\pm0.015$ \\% & $6.0~10^9$\\ 
CO(3-2) Blue & $-400\div -1000$ & DS3 & gauss & 12:56:14.24[$\pm0.09"$] & 56:52:25.00[$\pm0.08"$] & $26.0\pm5.8$ & $15.6\pm3.5$ & $0.6\pm0.4$\\% & $1.4~10^8$\\ 
CO(3-2) Red & $400\div 1000$ & DS3  & gauss & 12:56:14.23[$\pm0.05"$] & 56:52:25.22[$\pm0.05"$] & $51.6\pm6.0$& $31.0\pm3.6$ & $0.8\pm0.2$ \\% & $2.8~10^8$\\ 
%$^{13}$CS(5-4) & -700$\div$-1000$^*$ &  DS1 &  \\
          \hline
\end{tabular}
\end{center}
Note. - Errors: 
%for positions at CO(2-1), $\pm$; at CO(3-2), $\pm$;
flux errors are statistical, $\pm20\%$ should be accounted for flux calibration. 
The physical scale is 0.837 kpc/\arcsec. 
%*corresponding to a central observed frequency of $^{13}$CS(5-4) of the 221.986 
\label{tabvis}
\end{scriptsize}
\end{table*}

 \begin{figure*}[t]
\centering
\includegraphics[width=\textwidth]{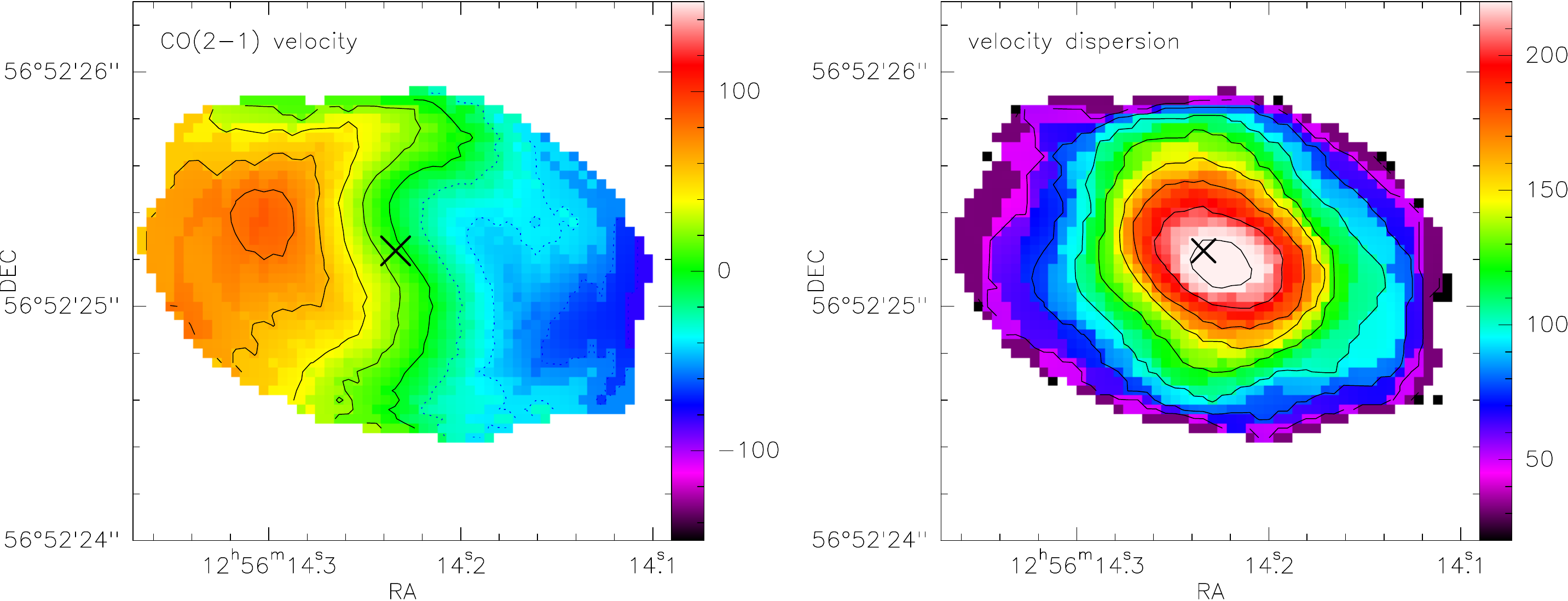}
\caption{Moment 1 map (left panel, levels are 20 km s$^{-1}$ each) and Moment 2 map (right panel, levels are 25 km s$^{-1}$ each) of the systemic component of CO(2-1). Cross: AGN position from Ma et al. (1998). 
}
\label{rota}
\end{figure*}

\begin{figure*}
\centering
\includegraphics[width=\textwidth]{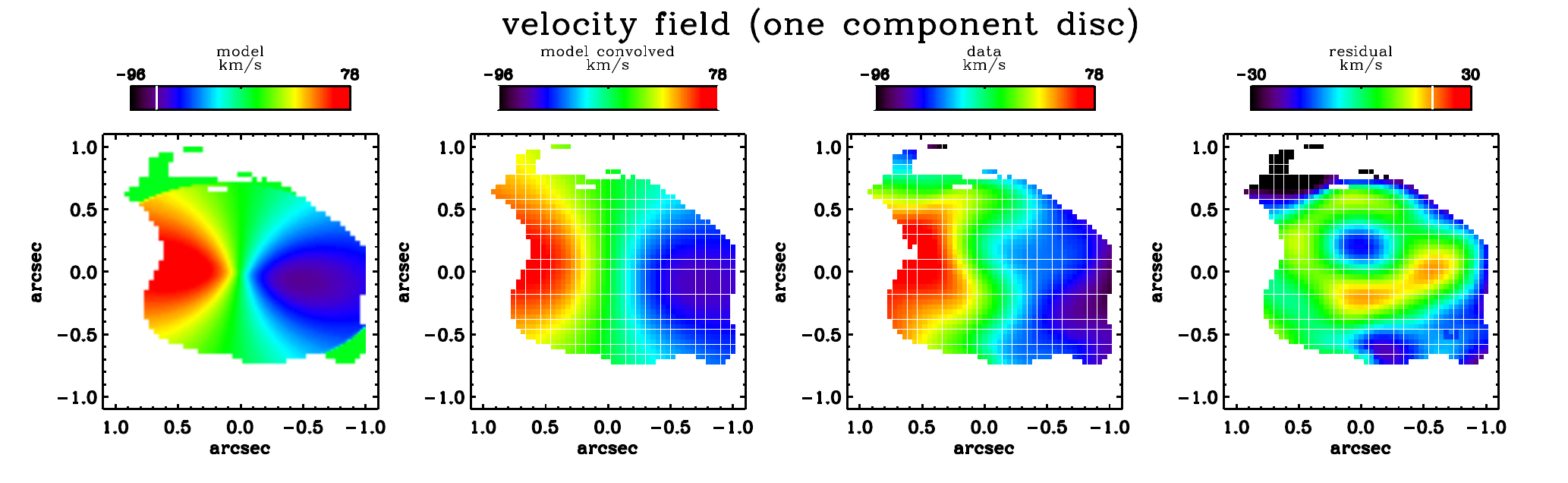}\\
\includegraphics[width=\textwidth]{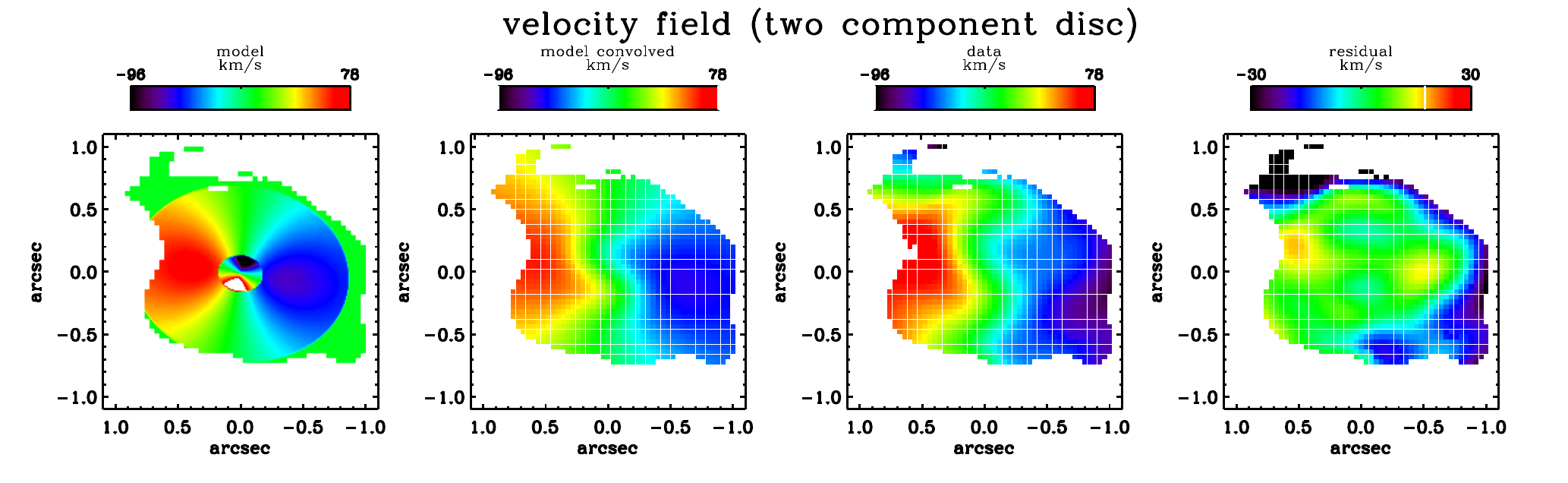}\\
\includegraphics[width=\textwidth]{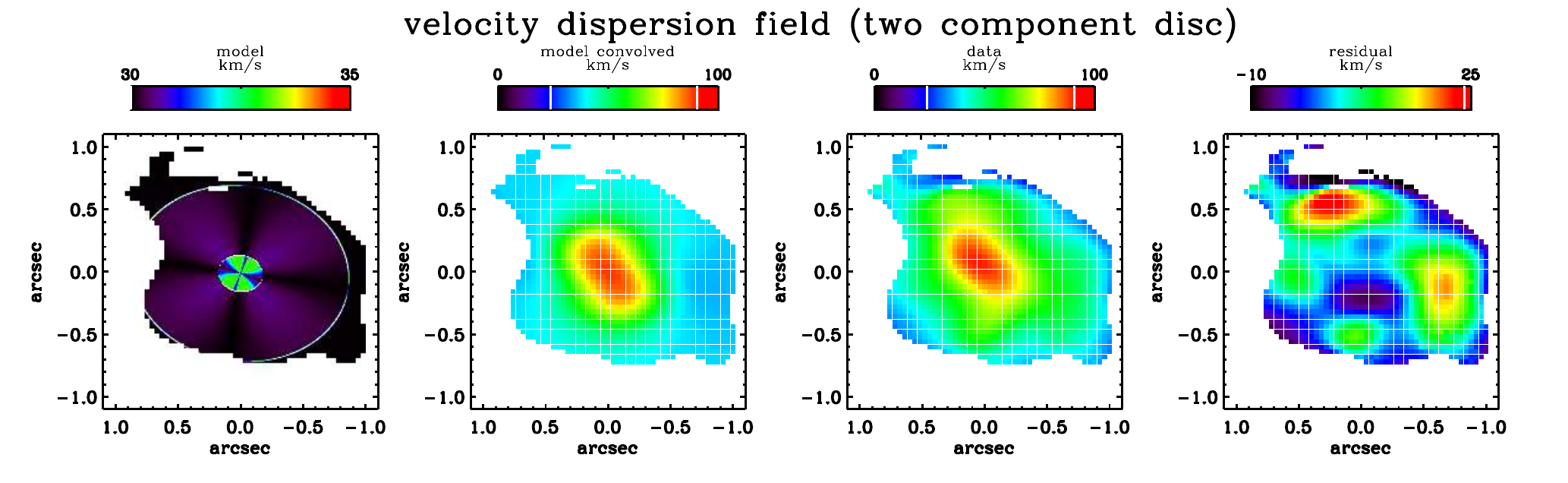}
\caption{From left to right: model, model folded with the beam, observed distribution and residuals. 
Top panels:  model of the molecular disk with an exponential disc function. 
Middle panels: model with two components, an outer ring and an inner exponential disk.  
 Bottom panels: velocity dispersion of the two-component model. 
 }
 \label{model}
\end{figure*}

%%Table disk parameters
\begin{table}[b]
\caption{Fit parameters derived for the molecular disk.}
%\begin{scriptsize}
\label{tabdisk}
\begin{center}
\begin{tabular}{lccc}
\hline
\multicolumn{1}{c} {Component}&
\multicolumn{1}{c} {$\rm r_0$}&
\multicolumn{1}{c} {PA}&
\multicolumn{1}{c} {Line-of-sight Inclination}\\
\multicolumn{1}{c} {}&
\multicolumn{1}{c} {[arcsec]}&
\multicolumn{1}{c} {[deg]}&
\multicolumn{1}{c} {[deg]}\\
\hline
Outer ring & 0.24 & $-12\pm5$ &  $36\pm10$\\
Inner disk  & 0.12  & $84\pm5$ &  $58\pm10$ \\
\hline
\end{tabular}
\end{center}
%\end{scriptsize}
\end{table}

\section{Results of (sub)-mm observations}

\subsection{Molecular disk}

We built moment 1 and 2 maps of the CO(2-1) emission using DS1. 
The Moment 1 (velocity) map of CO(2-1) in Figure \ref{rota}, left panel, shows that rotation occurs in approximately the east-west direction, as previously measured  by Downes \& Solomon (1998), and the molecular disk extends out to a radius about 800 pc. 
The S-shaped pattern (Fig. \ref{rota}) indicates that the kinematics in the inner disk deviates from a simple east-west rotation, and suggests the presence of an inner warped disk on scales of $\leq0.2\arcsec$, tilted with respect to the main rotation plane. 
%Fainter emission from rotation is seen on a total extent of 2.5$\arcsec$, corresponding to a outer disk radius of $\sim 1$ kpc (Figure \ref{pvds1}).
The Moment 2 map (Figure \ref{rota}, right panel) is peaked close to the AGN, and shows branches of emission in the north, south and west directions, also indicating an additional kinematical component. 
%%%%here map of the velocity dispersion copy from Aalto 

By fitting an exponential disk model to the visibilities in the velocity range -300 to 300 km s$^{-1}$ (systemic component) we derive a similar  size of the molecular disk of 720 pc FWHM (Table 4).  The disk size derived from CO (3-2)  is about 900 pc.
To better constrain the properties of the molecular disk, we first fit the flux and velocity maps of the molecular disk (systemic component) using the kinematical disk model by Gnerucci et al. (2011a). We assumed that 
(i) the gas is circularly rotating in a thin disk, (ii) the gravitational potential depends only on the disk mass, and (iii) the disk surface mass density distribution is exponential, i.e.

\begin{equation}
\Sigma(r) = \Sigma_0~ e^{-r/r_0}
\end{equation}

\noindent
where $r$ is the distance from the center and r$_0$ is the scale radius.  
Following Cresci et al. (2009), the adopted surface brightness profile for both components is exponential

\begin{equation}
I(r) = I_0~ e^{-r/r_0}
\end{equation}

The disk model and fit results are shown in Fig. \ref{model}, top panel.  
The model is then convolved with the beam of the data. The flux and velocity maps are fitted with a minimizing-$\chi^2$ method.
Residuals due to the inner S-shaped feature are seen in the velocity map (Fig. \ref{model}, top right panel). 

In order to account for the inner tilted disk we fit the velocity map with two components, i.e.,  an inner disk and an outer ring with the same  exponential profile but different scale radii $r_0$, and allowing for different inclinations ($i$) and position angles (PA). Position angles are measured in deg and are positive from east via north. 
This fit gives for the outer ring a scale radius of  r$_0=0.24$ arcsec, a PA of the line of nodes of $ -12 \deg$, and inclination of $i=36\pm 10 \deg$ (including a dominant contribution from systematic uncertainties). 
For the inner disk, we find a scale radius of $r_0=0.12$, a PA$ = 84\deg$, inclination $= 58\pm10 \deg$ (Table \ref{tabdisk}).
The size (radius 200 pc), inclination and orientation of the velocity gradient (nearly north-south) of the inner warped disk of CO(2-1) match well those seen in the vibrationally excited transition HCN(3-2) $v_2=1$ (Aalto et al. 2014), and that seen in the OH megamaser observation of Klockner et al. (2003).

The dynamical mass enclosed within 400 pc from the AGN is $1.6^{+1.1}_{-0.5} \times 10^{9}~M_{\odot}$ (with the errors dominated by the uncertainty on the inclination). 
Downes \& Solomon (1998) found $\rm 1.27\times 10^{10}~M_{\odot}$ for an inclination of 10 deg (based on the ratio of the two semi-axis of the disk, and, therefore, probably affected by even larger uncertainties),  
with our best fit inclination of 36 deg it would be a factor of 10 smaller and, thus, in agreement with our measurements.
%Conversely, Hinz \& Rieke (2006) reported a dynamical mass of $10^{10}~M_{\odot}$  from Pisces H and K band. [QUESTO DISCREPA, LEGGERE BENE IL PAPER E COMMENTARE]

\begin{figure*}
\centering
\includegraphics[scale=0.2]{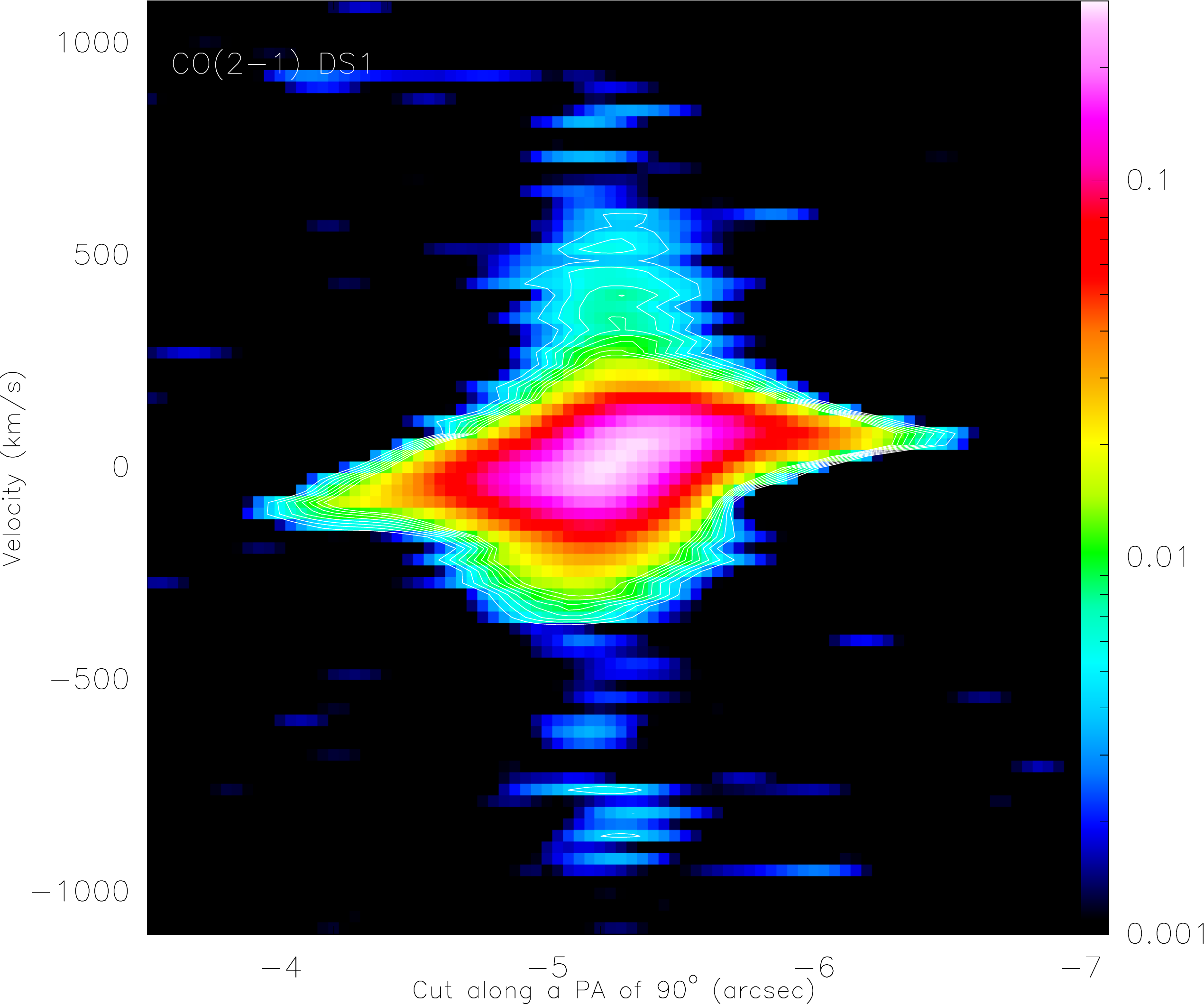}
\includegraphics[scale=0.2]{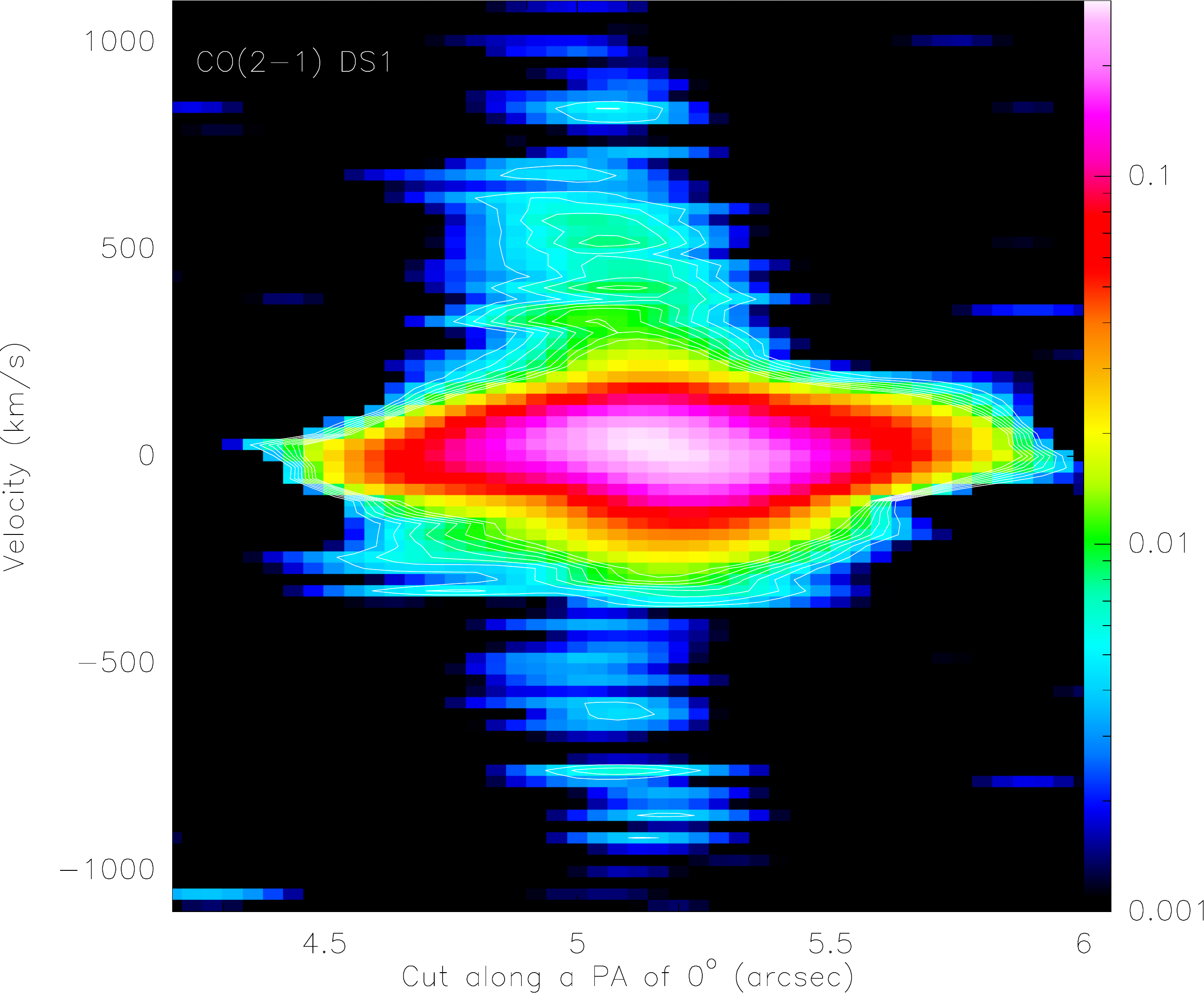}
\includegraphics[scale=0.2]{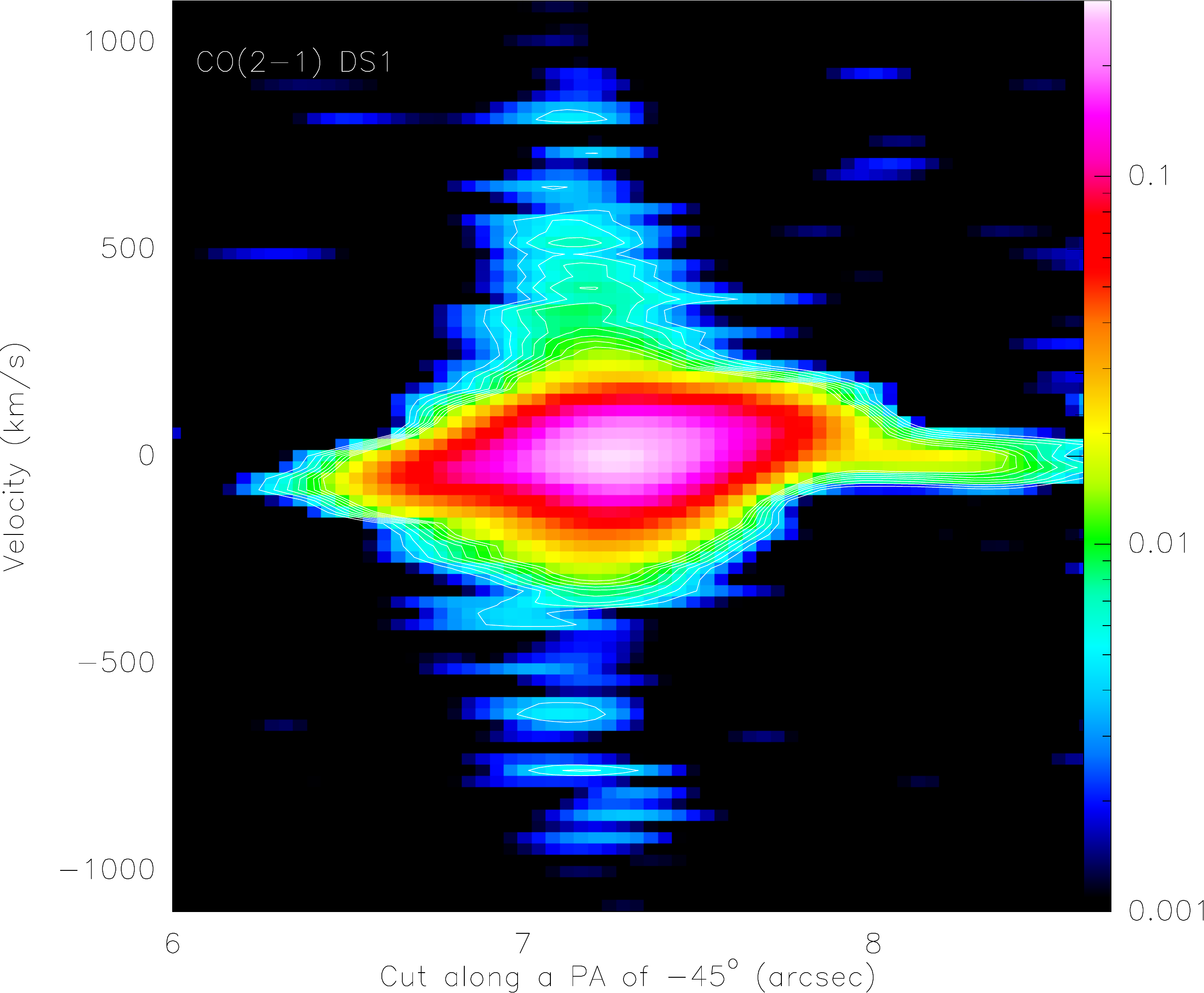}
\caption{Position-velocity plots with east-west (left panel, left to right), south-north (middle panel, left to right) and PA -45 deg, south-west to north-east (right panel, left to right ) cuts, through the CO(2-1) peak from DS1.  
Contours levels are 3 to 15$\sigma$, 1$\sigma=$1.3 mJy/20 MHz.
 }
\label{pvds1}
\end{figure*}

\begin{figure*}
\centering
\includegraphics[scale=0.205]{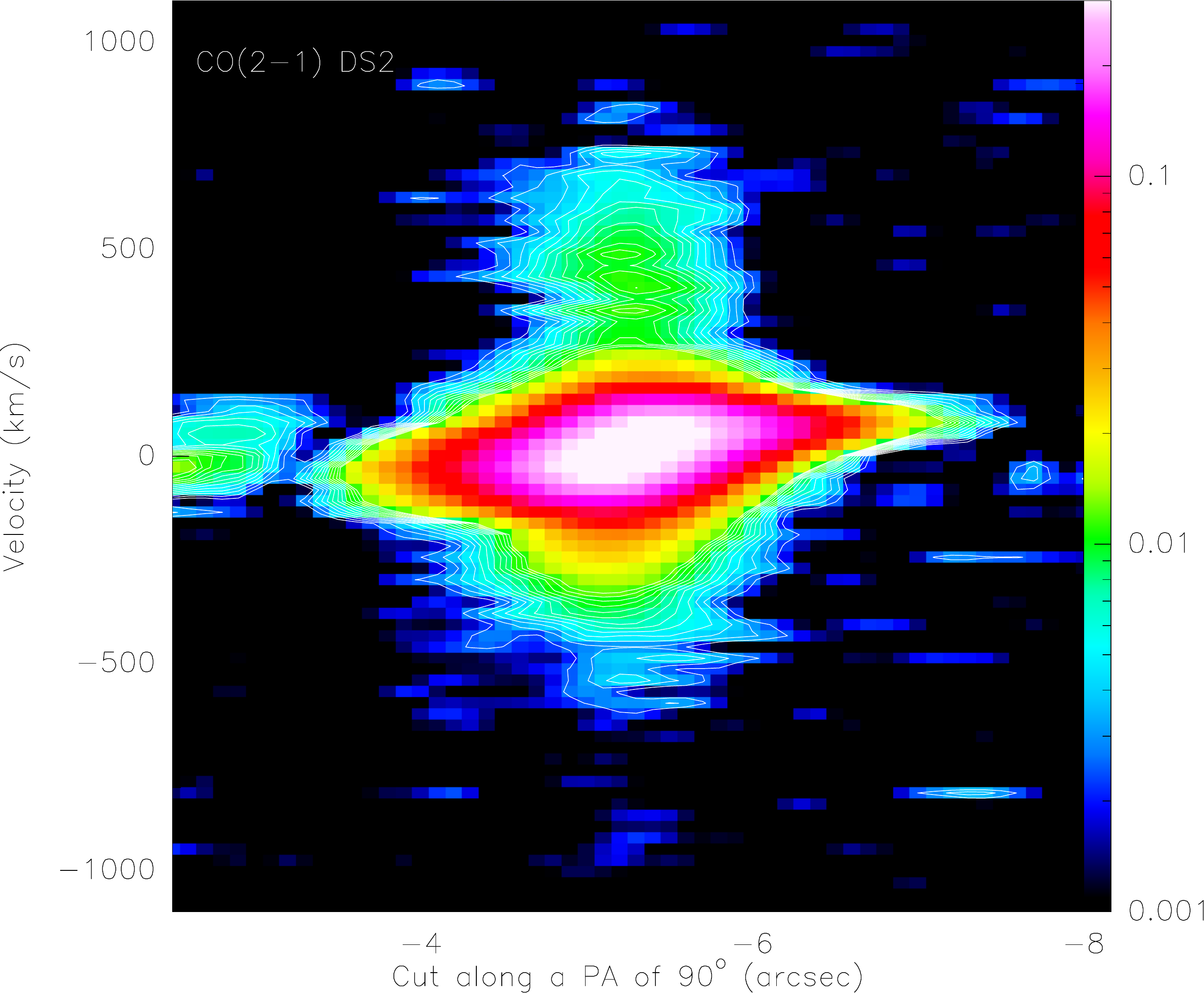}
\includegraphics[scale=0.205]{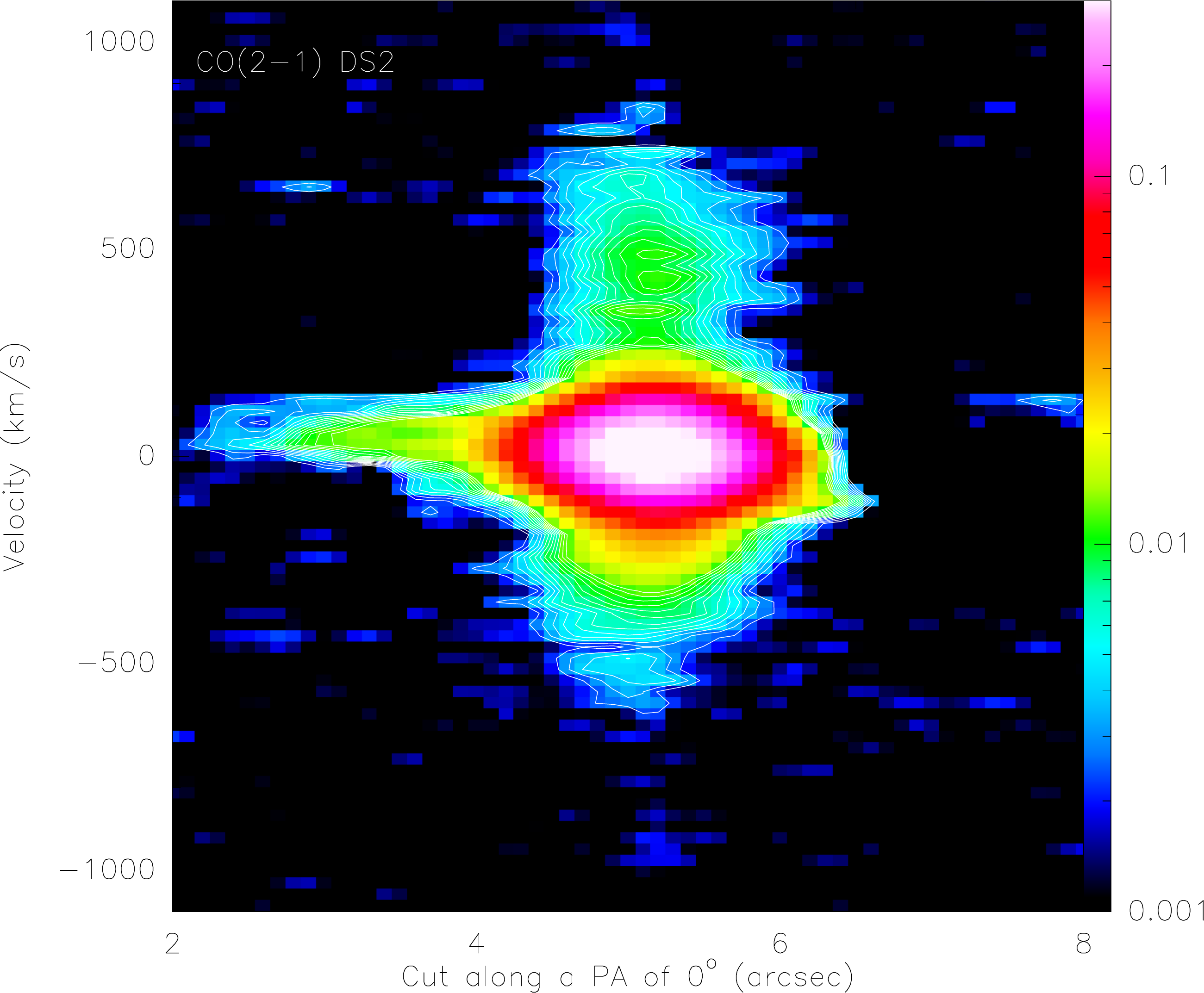}
\includegraphics[scale=0.205]{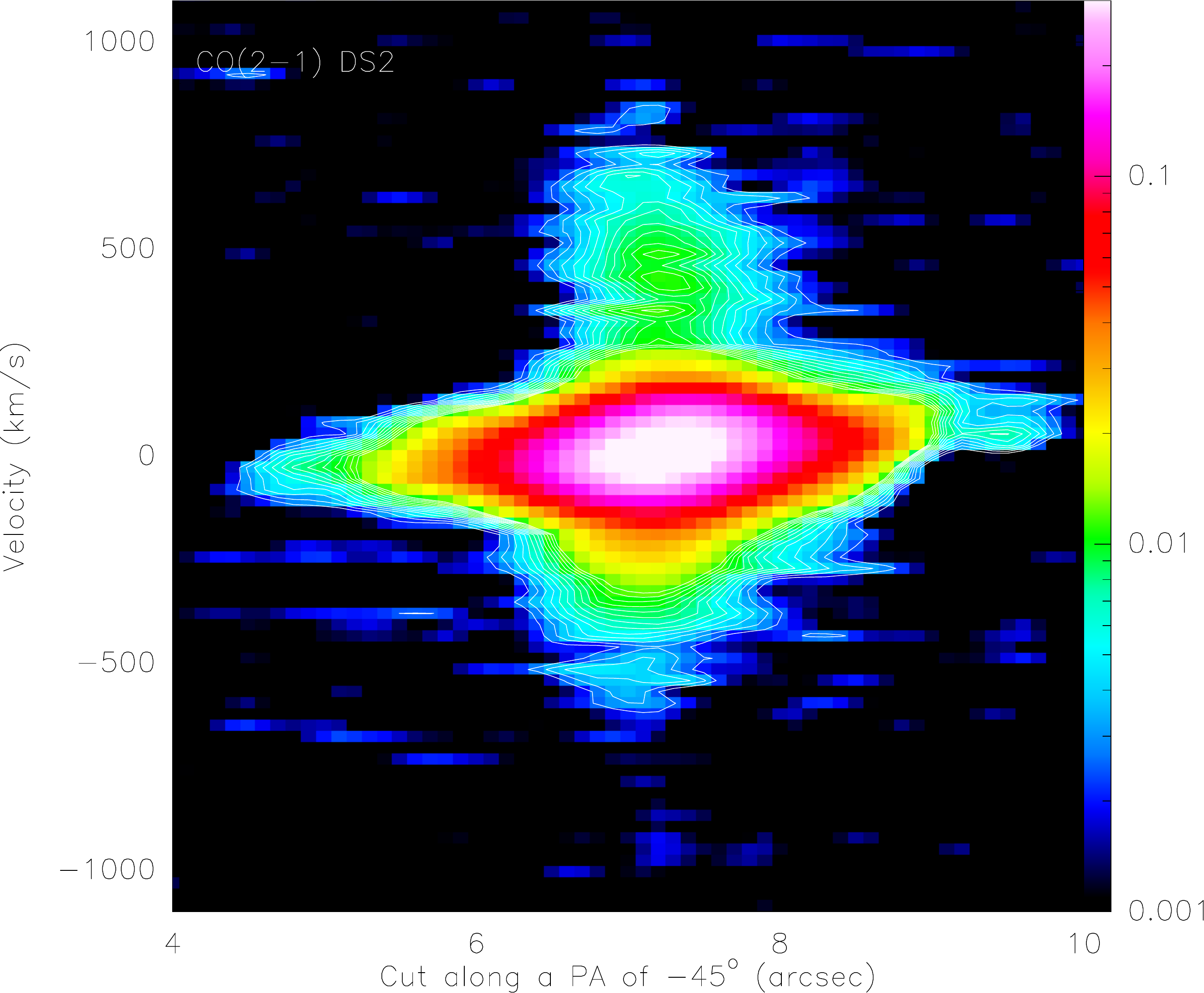}
\caption{Position-velocity plots of CO(2-1) from DS2 along the same slices than in Fig. \ref{pvds1}. 
 Contours levels are 3 to 15$\sigma$, 1$\sigma=$0.8 mJy/20 MHz.
 }
\label{pvds2}
\end{figure*}

\begin{figure*}
\centering
\includegraphics[scale=0.205]{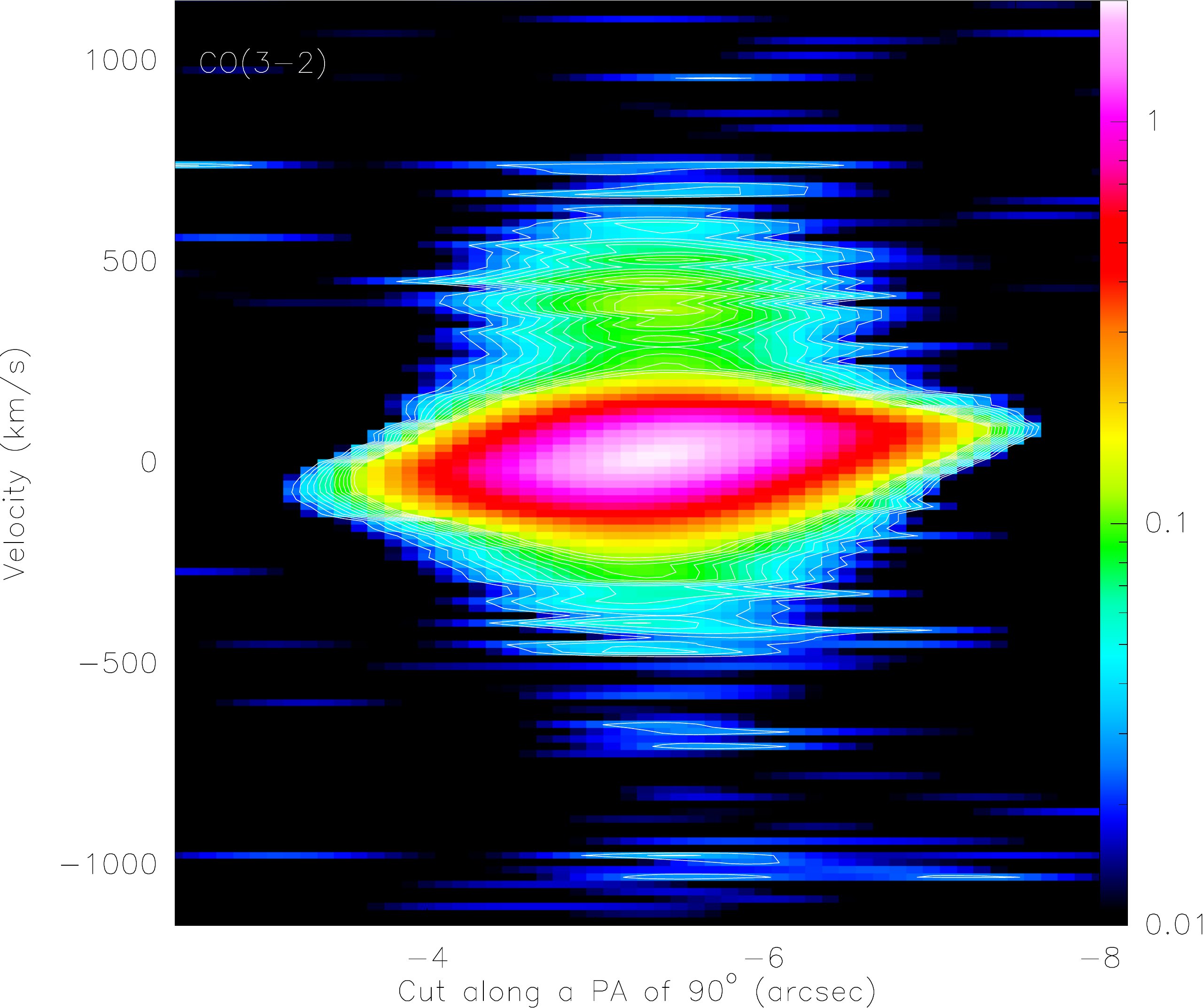}
\includegraphics[scale=0.205]{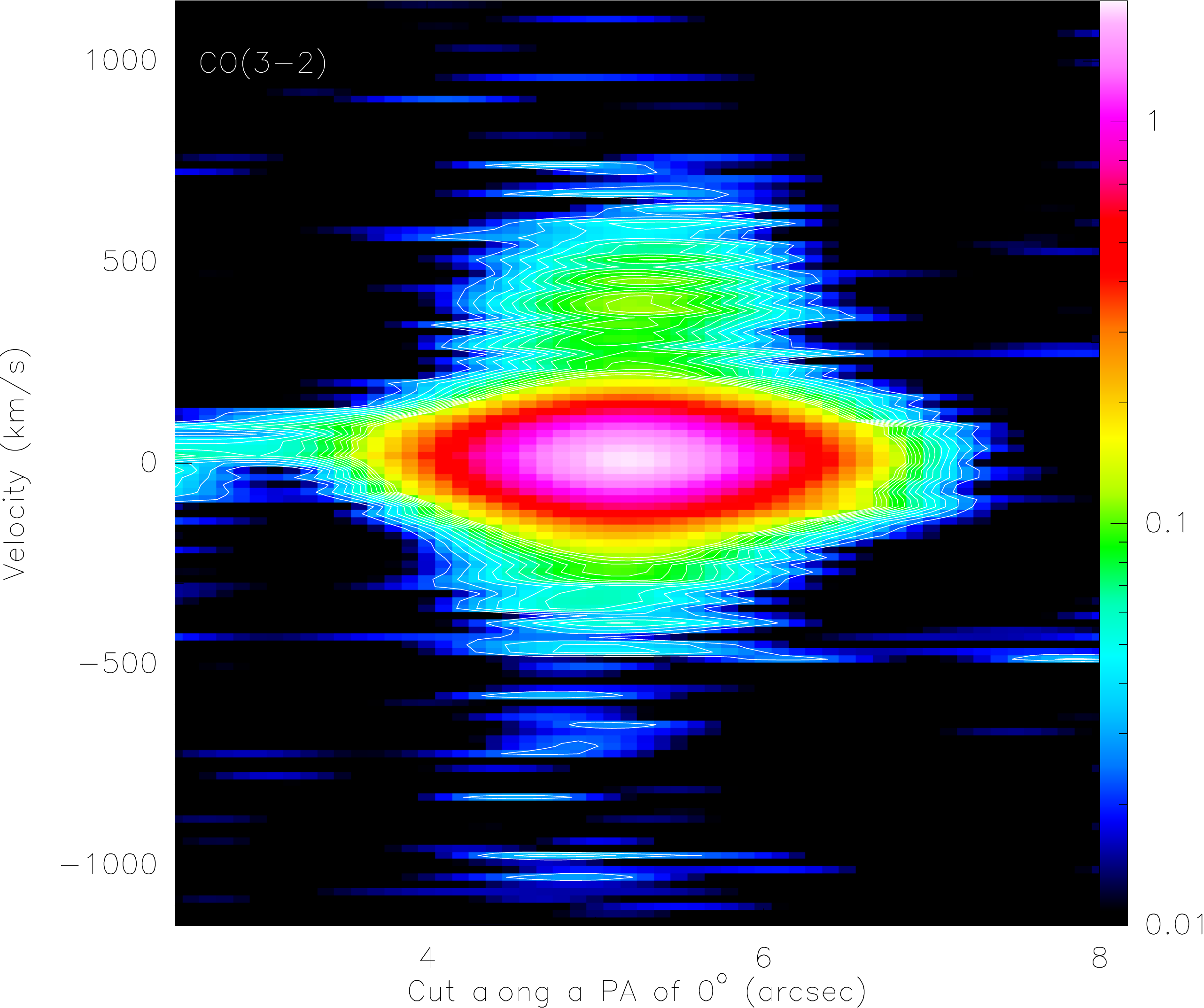}
\includegraphics[scale=0.205]{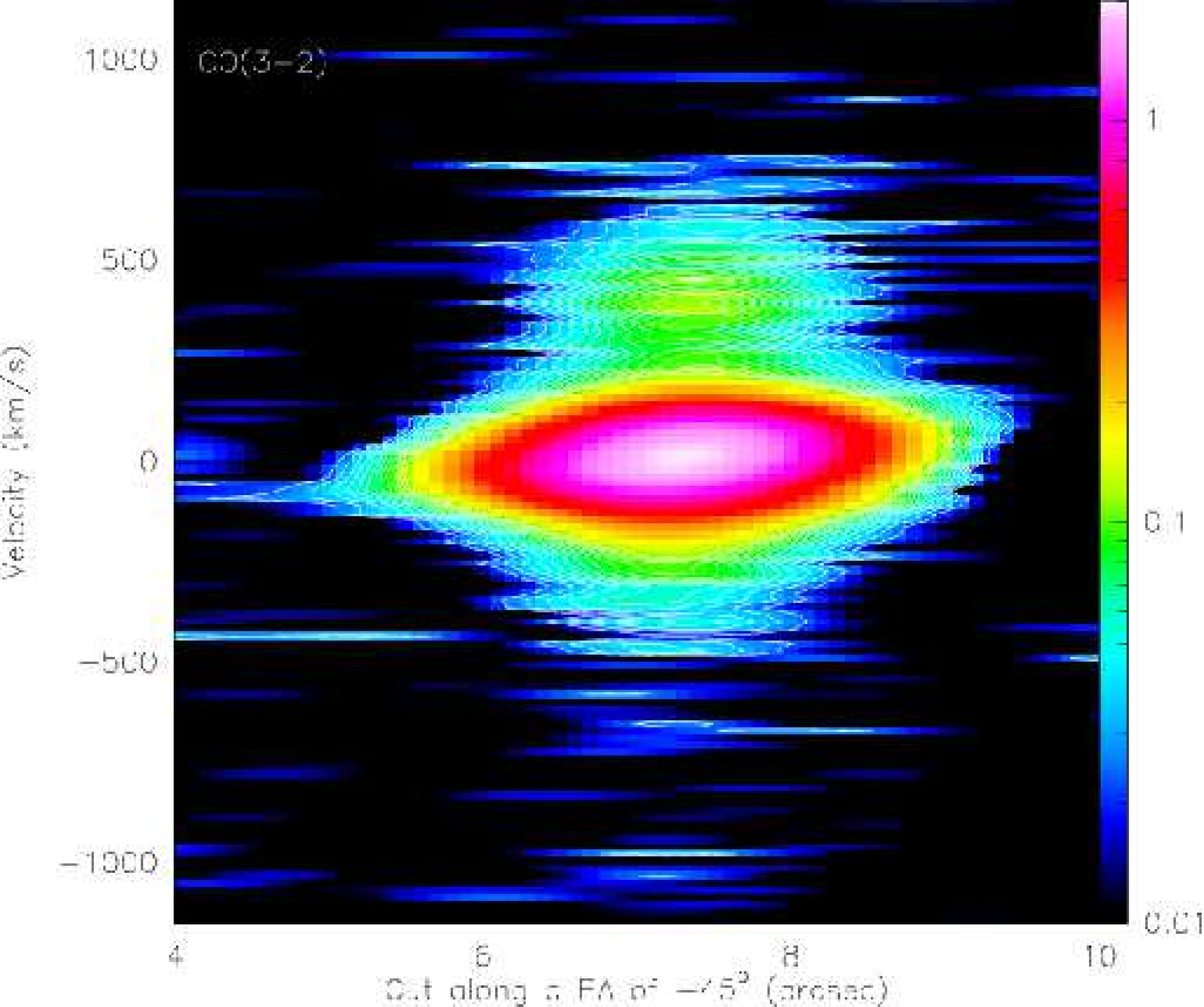}
\caption{Position-velocity plots of CO(3-2) along the same slices than in Fig. \ref{pvds1}. 
 Contours levels are 3 to 15$\sigma$, 1$\sigma=$0.8 mJy/20 MHz.
 }
\label{pvds3}
\end{figure*}

\subsection{Molecular outflow}

\begin{figure*}
\centering
\includegraphics[width=\textwidth]{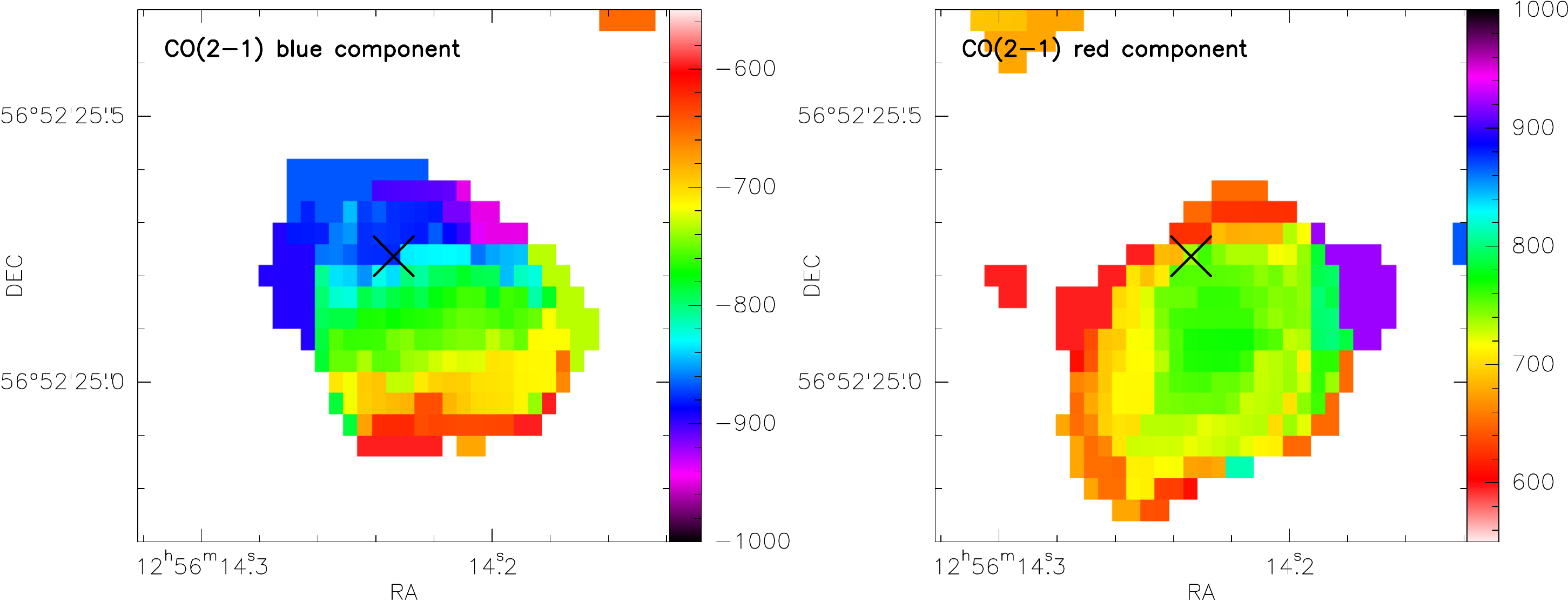}
\caption{Velocity maps of the blue (left) and redshifted CO emission (right).  The velocity (in km s$^{-1}$ with respect to the systemic velocity, i.e. the CO(2-1) peak, is shown in the colour bars. The cross marks the AGN VLBI position.}
 \label{velo-map}
\end{figure*}

Table \ref{tabvis} reports the results of the visibility fits of CO (2-1) and (3-2). 
The FWHM size of the high velocity wings of CO(2-1) is 0.35$\arcsec$ (0.3 kpc) and 0.44$\arcsec$ (0.4 kpc) for the red and blue components, respectively. 
%In  the clean maps, in addition to the compact emission fitted by the models, we measure extended emission with lower surface brightness with spatial extension of 0.7-1 kpc (Figure \ref{channel-wings}),  consistent with what estimated with shorter baseline data (Cicone et al. 2012), and which cannot be fitted by a simple model. 
In addition we detect lower surface brightness extended emission due to receding (redshifted) gas out to at least 1 kpc north-east off the AGN (Fig. 3).
%In this region the AGN jet is thought to give an extra acceleration to the outflow, therefore creating a pile up of emission, as also suggested by the neutral outflow (Rupke \& Veilleux 2013).  
The CO (3-2) outflow component has a FWHM size of $0.6$ (500 pc) and 0.8$\arcsec$ (700 pc) for the blue and red components, respectively. 

Position-velocity plots along three different directions are shown in Figures \ref{pvds1} to \ref{pvds3}.  
For each data set, we extracted a horizontal (i.e. along the major axis of disk rotation, west-east cut), a vertical (south-north)  and a $-45$ deg (south-west to north-east) cut. 
In all three data sets the rotation pattern of the outer disk is seen in the west-east cuts.  Rotation is also visible along the other two directions in the high resolution data.  
Emission with velocity close to the systemic ($\pm300$ kms/s) is extended out to about 1 arcsec from the nucleus.
%2\arcsec~ west and 2.5\arcsec~ south of the peak emission, similarly to what seen in HCN(3-2) (Aalto et al. 21014, 2012).

The high velocity gas ($v>400$ km s$^{-1}$) is seen in all data sets,
and along all the examined directions, suggesting that the velocity
distribution is rather isotropic.  However, both the high speed
receding and approaching gas components extend southward of the
nucleus, with the approaching one further elongated along the
south-west direction (see middle and right panels of Figure 6).

Figure \ref{velo-map} shows the Moment 1 maps of the red and blue
CO(2-1) wings in the velocity ranges -500 to -1000 and 500 to 900 km
s$^{-1}$, respectively.  In the blue component a velocity gradient is
seen. The speed of the gas is $\sim-900$ km s$^{-1}$ at the position
of the nucleus, and decreases to $-600$ km s$^{-1}$ at $0.4\arcsec$
south-west of the nucleus. This is consistent with the
position-velocity plot along the diagonal cut (Fig. \ref{pvds1}, right
panel).  The bulk of the gas receding at velocities of 700-800 km
s$^{-1}$ is also located at $\sim0.3 $ arcsec south-west of the AGN
(Figure \ref{velo-map}, right panel), and shows a milder velocity
gradient along approximately the same axis as the approaching gas.
Emission with velocity of about 700 to 800 km s$^{-1}$ is also seen at
(+0.8,+0.8$\arcsec$) offset from the phase center, along the same
diagonal direction where the blueshifted velocity gradient is
found. We cannot exclude that this feature is a residual side lobe,
but the absence of the corresponding side lobe in the opposite
direction, suggests that this feature might be real.

\subsection{Mapping $\dot M_{OF}$ and $\dot E_{kin,OF}$ }

\begin{figure}
\centering
\includegraphics[width=6cm]{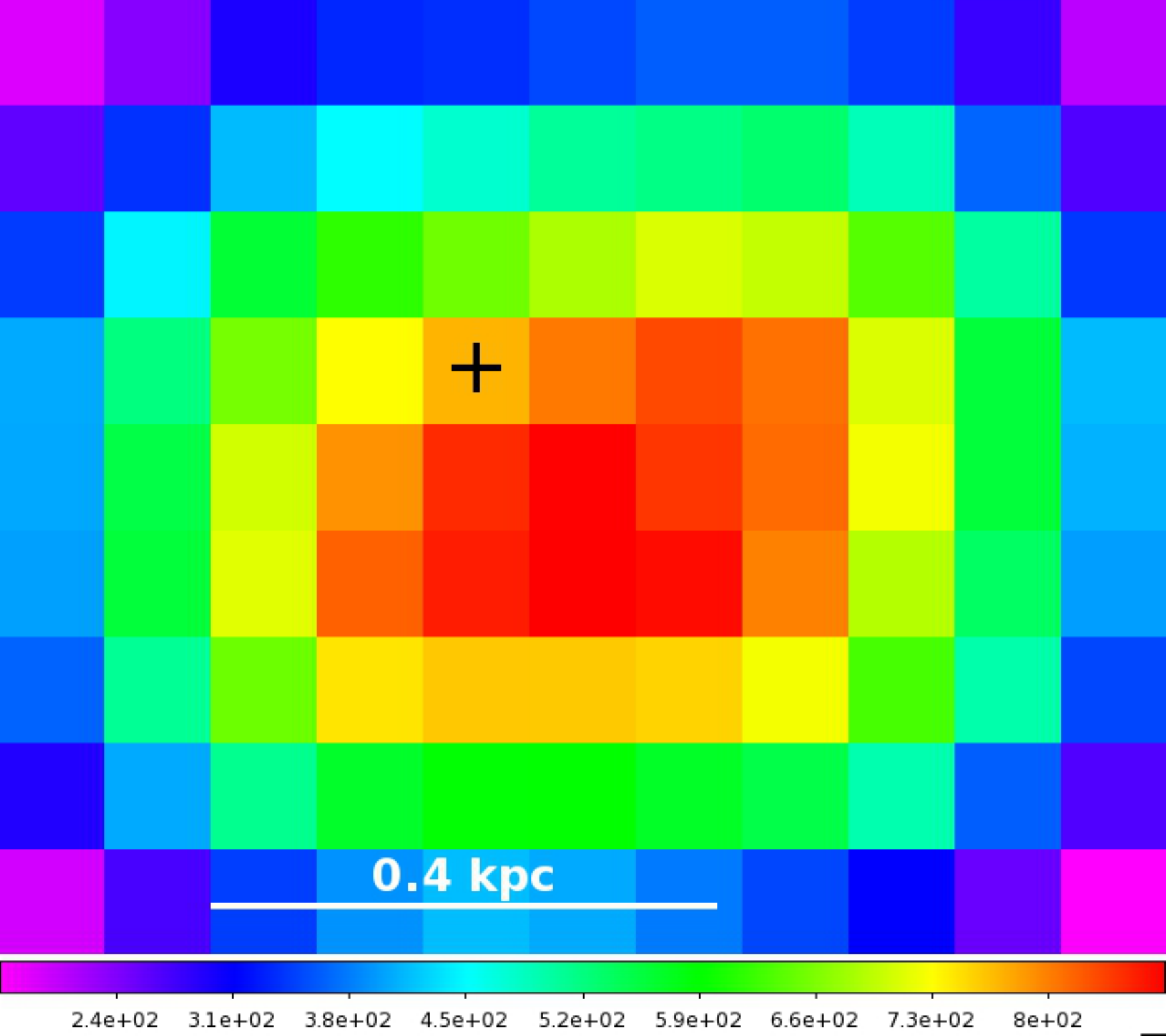}
\caption{Maps of $\rm v_{max}=velocity~shift_{broad}+2\sigma_{broad}$ (colour scale in km s$^{-1}$). 
The pixel size is 0.1 arcsec. The cross marks the AGN VLBI position.
}
 \label{vmax}
\end{figure}

\begin{figure}
\centering
\includegraphics[width=6cm]{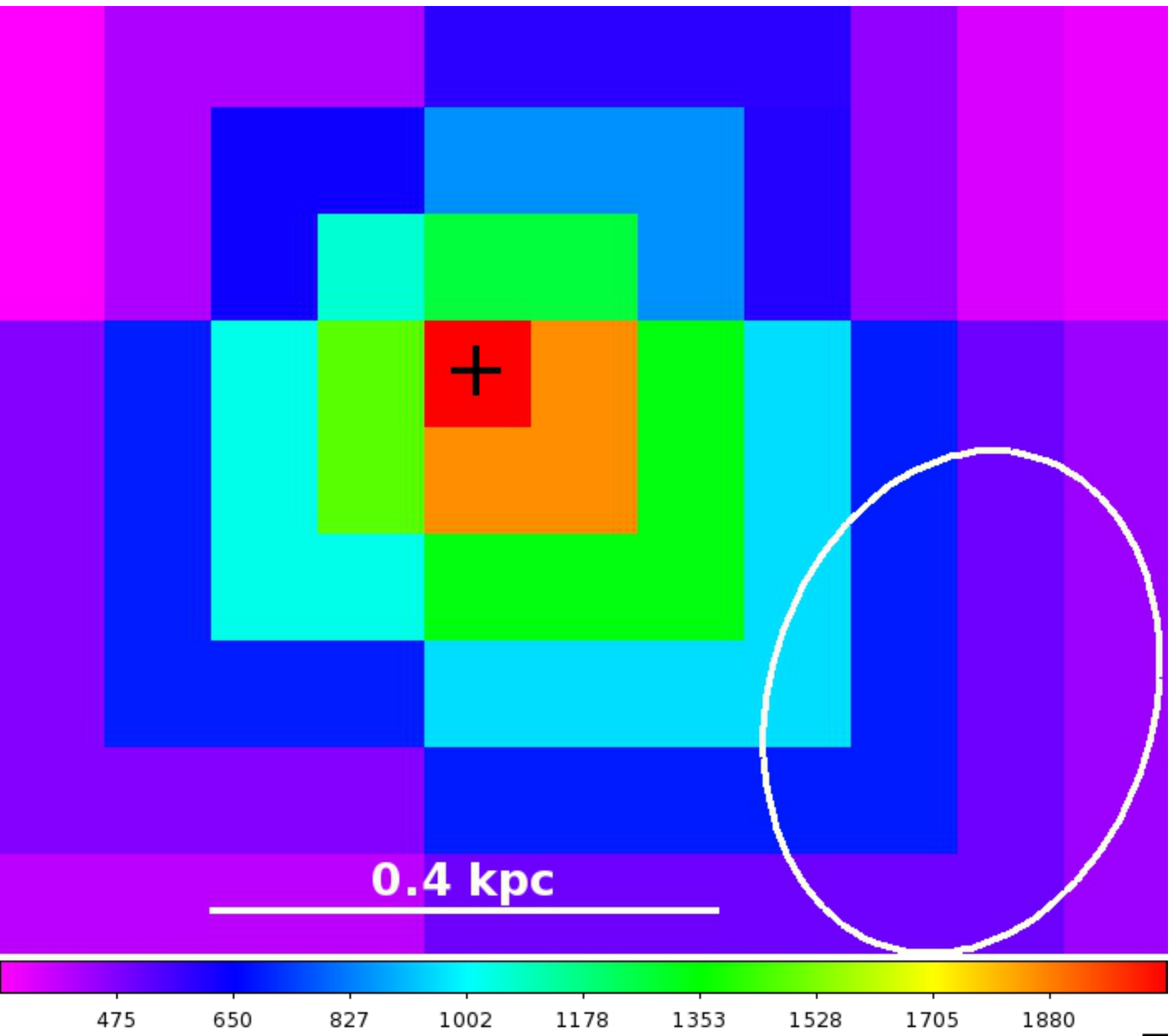}
\caption{Map of $\rm \dot M_{OF}$ (the scale in $\rm
  M_{\odot}~yr^{-1}$ is reported at the panel bottom).  M$_{OF}$
    is computed by averaging the CO flux of the broad gaussian
    components in quadrants of increasing radii (south-west,
    south-east, north-east, north-west squared sectors off the AGN),
    and then converting to M(H2) using a conversion factor of 0.5.
  The clean beam is shown as an open ellipse. The physical scale is
  0.1 arcsec/pixel. The cross marks the AGN VLBI position.}
 \label{rmout-maps}
\end{figure}

We fitted the pixel-by-pixel spectra with three Gaussian functions, in
order to simultaneously account for both low velocity (rotating disk)
and high velocity gas (outflows).  We define
the maximum velocity, $\rm v_{max}$,  as the shift between the velocity peak of
broad component and the systemic velocity plus 2$\sigma$, 
where $\sigma$ is the velocity dispersion of the broad gaussian component 
($ v_{max}=velocity
~shift_{broad}+2\sigma_{broad}$, $\sigma_{broad}=FWHM_{broad}/2.35$). 
The $\rm v_{max}$ map is show in Figure \ref{vmax}. 
Based on these three component fits, we attempted
to map the mass outflow rate and kinetic energy.  In previous works
(e.g. Feruglio et al. 2010) the mass outflow rate has been computed
using the continuity fluid equation

\begin{equation}
\rm \dot M_{OF}= \Omega ~ R_{OF}^2 ~\rho_{OF}~ v_{max}
\label{fluid}
\end{equation}

where $\rho_{OF}$ is the average mass density of the outflow,
$v_{max}$ is the maximum velocity of the outflowing gas, and R$_{OF}$
is the radius at which the outflow rate is computed, and $\Omega$ is
the solid angle subtended by the outflow.  Assuming a spherical
sector, $\rho_{OF}= 3 M_{OF} / \Omega R_{OF}^{3}$, then, $\dot
M_{OF}=3\times v_{max}\times M_{OF}/R_{OF}$.  In this formula,
R$_{OF}$ is the distance from the nucleus (RA,DEC= 12:56:14.23,
56:52:25.20). Accordingly, $\rm \dot M_{OF}$ represents the instantaneous
outflow rate of the material at the edge R$_{OF}$ (i.e., it is a local
estimate) and it is three times larger than the total outflow mass
divided by the time required to push this mass through a spherical
surface of radius R$_{OF}$.  This estimator does not depend on the
solid angle $\Omega$ subtended by the outflow. 
Figure  \ref{rmout-maps} is the map of the outflow rate $\rm \dot
  M_{OF}$ derived by averaging the CO flux per beam of the broad
  gaussian components in quadrants of increasing radii (south-west,
  south-east, north-east, north-west off the AGN), and by adopting
  $\alpha_{CO}=0.5$ to convert from line luminosity to gas mass (Weiss et
    al. 2001). 
    $\rm \dot M_{OF}$ in Figure \ref{rmout-maps}, therefore,
    should not be considered a {\it pixel by pixel} estimator of the
    mass outflow rate, but rather the value in square sectors at
    increasing radii.

The outflow originates from the nucleus, and $\rm \dot M_{OF}$
decreases in all directions with increasing distance from the nucleus.
The outflow expands preferentially toward south-west, out to a distance of 0.4
arcsec from the nucleus.

\subsection{$\dot M_{OF}$, $\dot E_{kin,OF}$ and $M_{OF}/M_{disk}$ radial profiles}

In the following  we study the
mass outflow rate, kinetic power and  $\rm M_{OF}/M_{disk}$ as a function of the distance from the nucleus. 
Since the outflow is
located near the nucleus and rather symmetric, for the sake of
simplicity we adopted an one dimensional average.
We used DS1 to map the region at $<0.5$ kpc with the
best angular resolution, and DS2 to constrain any fainter high speed
emitting gas further out from the nucleus.   
We extracted spectra from box regions centered on the nucleus and with
increasing sides. We then fitted the spectra with three Gaussian
functions (or four when the systemic line needs two components) to
model both the rotating disk and the outflow components, and computed
the mass outflow rate as detailed above. The gas masses of the disks
(traced by the systemic CO line) and outflow (broad CO components)
have been calculated by integrating the best fit models and by using
$\alpha_{CO}=0.5$.

\begin{figure}
\centering
\includegraphics[width=\columnwidth]{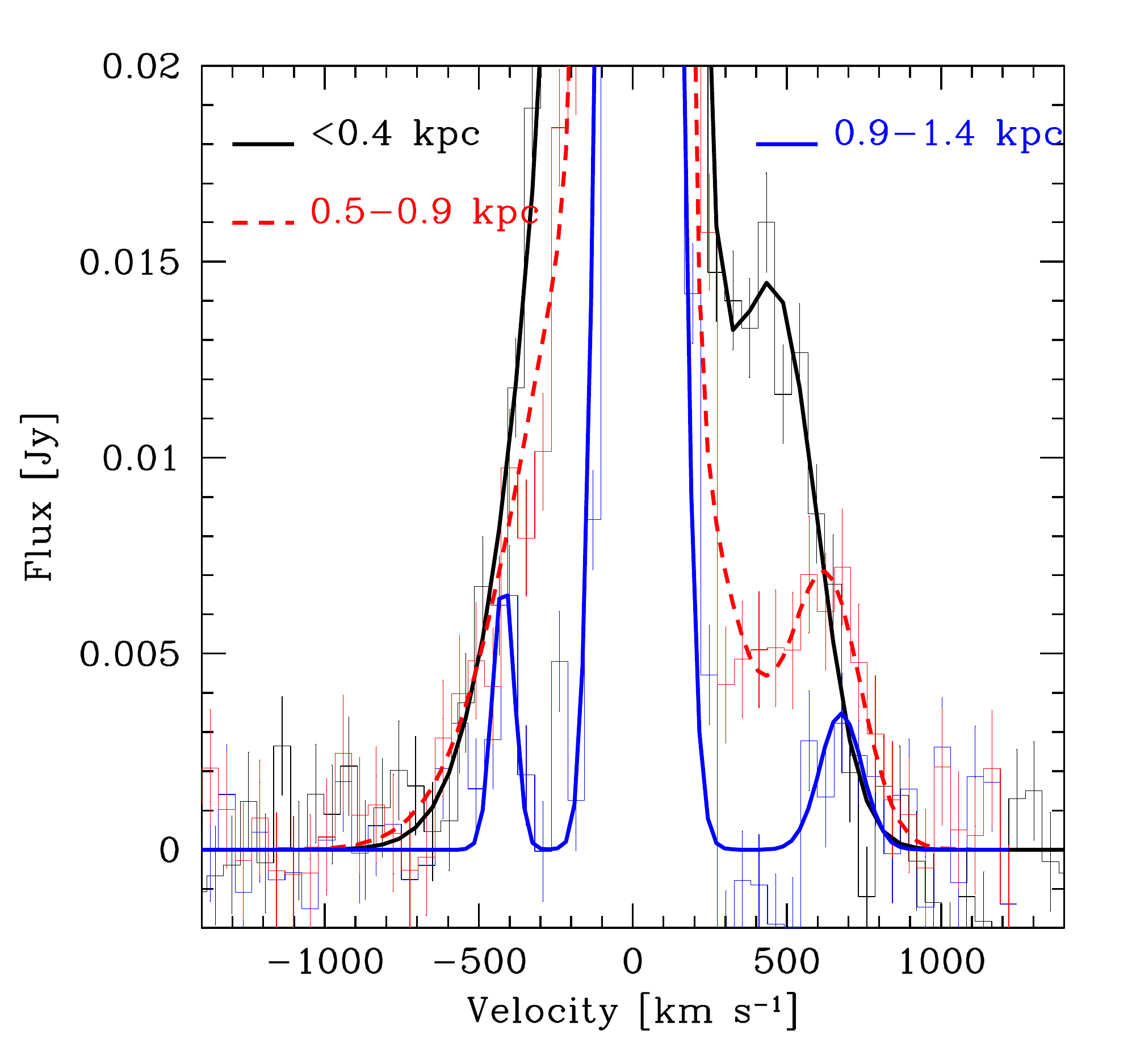}
\caption{ CO(2-1) emission line profiles extracted from square regions at different distances from the nucleus, as indicated by the colour-coded labels, and their multi-Gaussian best fit.}
 \label{specs}
\end{figure}

\begin{figure*}
\centering
\includegraphics[scale=0.31]{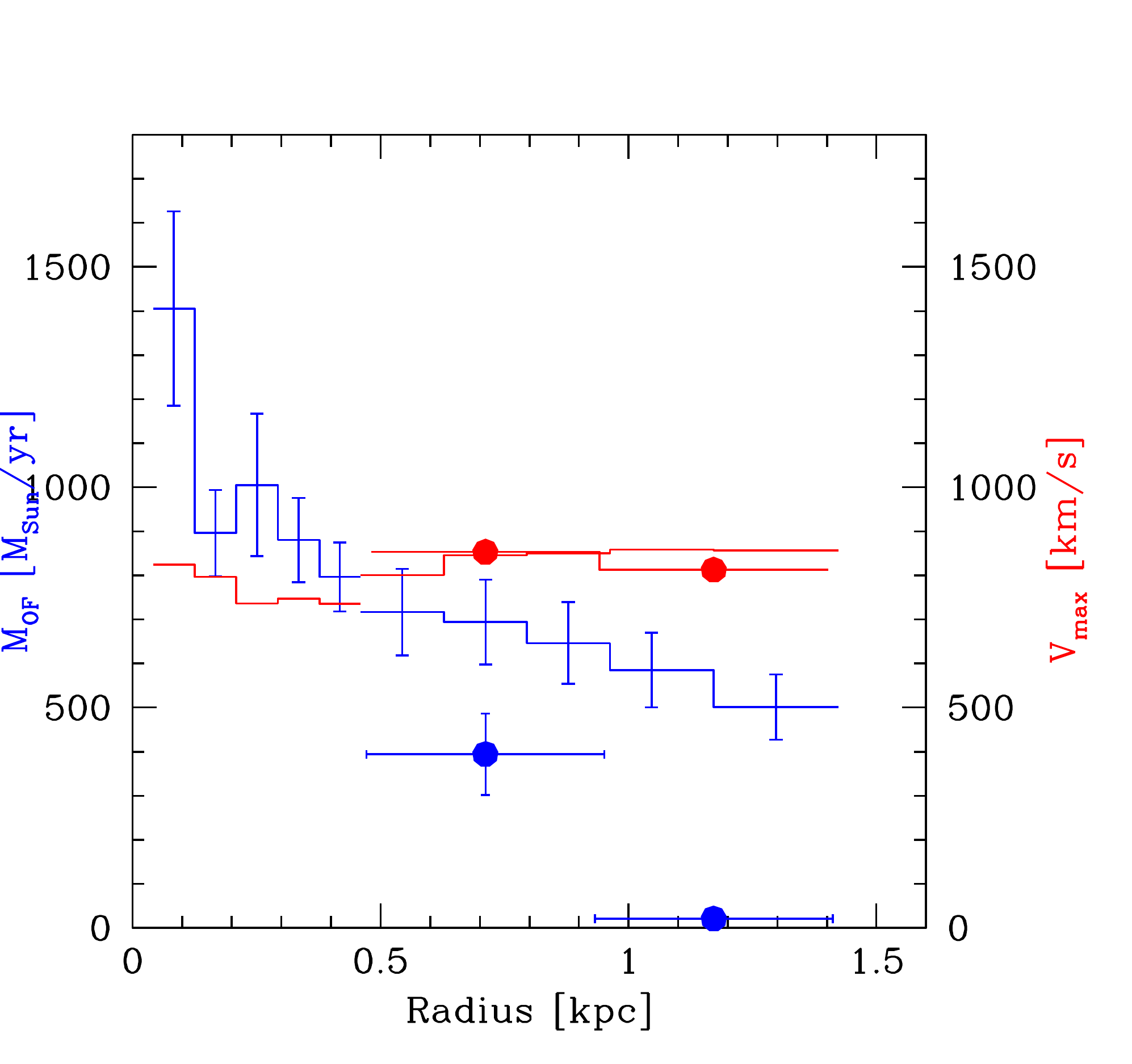}
\includegraphics[scale=0.29]{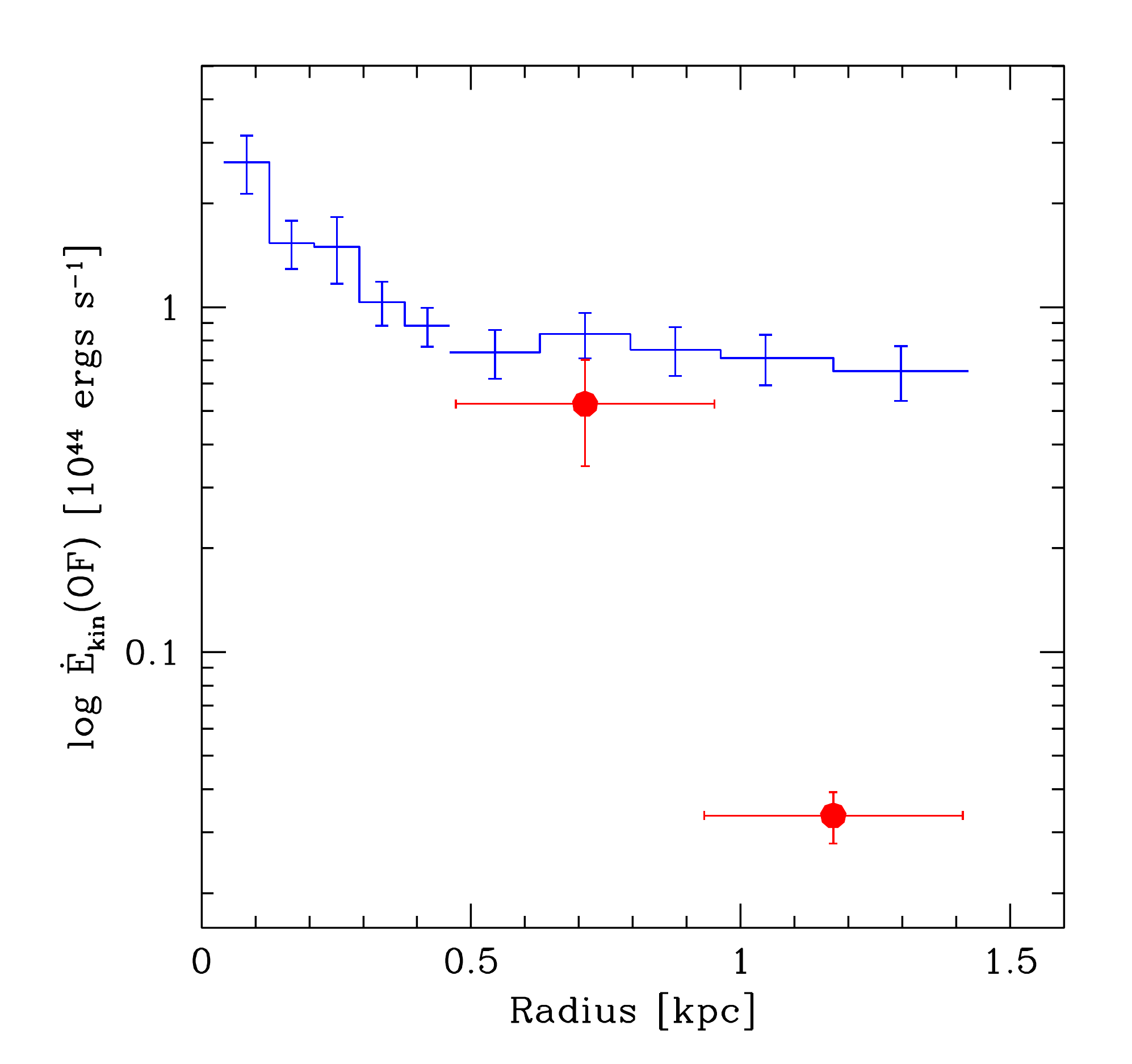}
\includegraphics[scale=0.29]{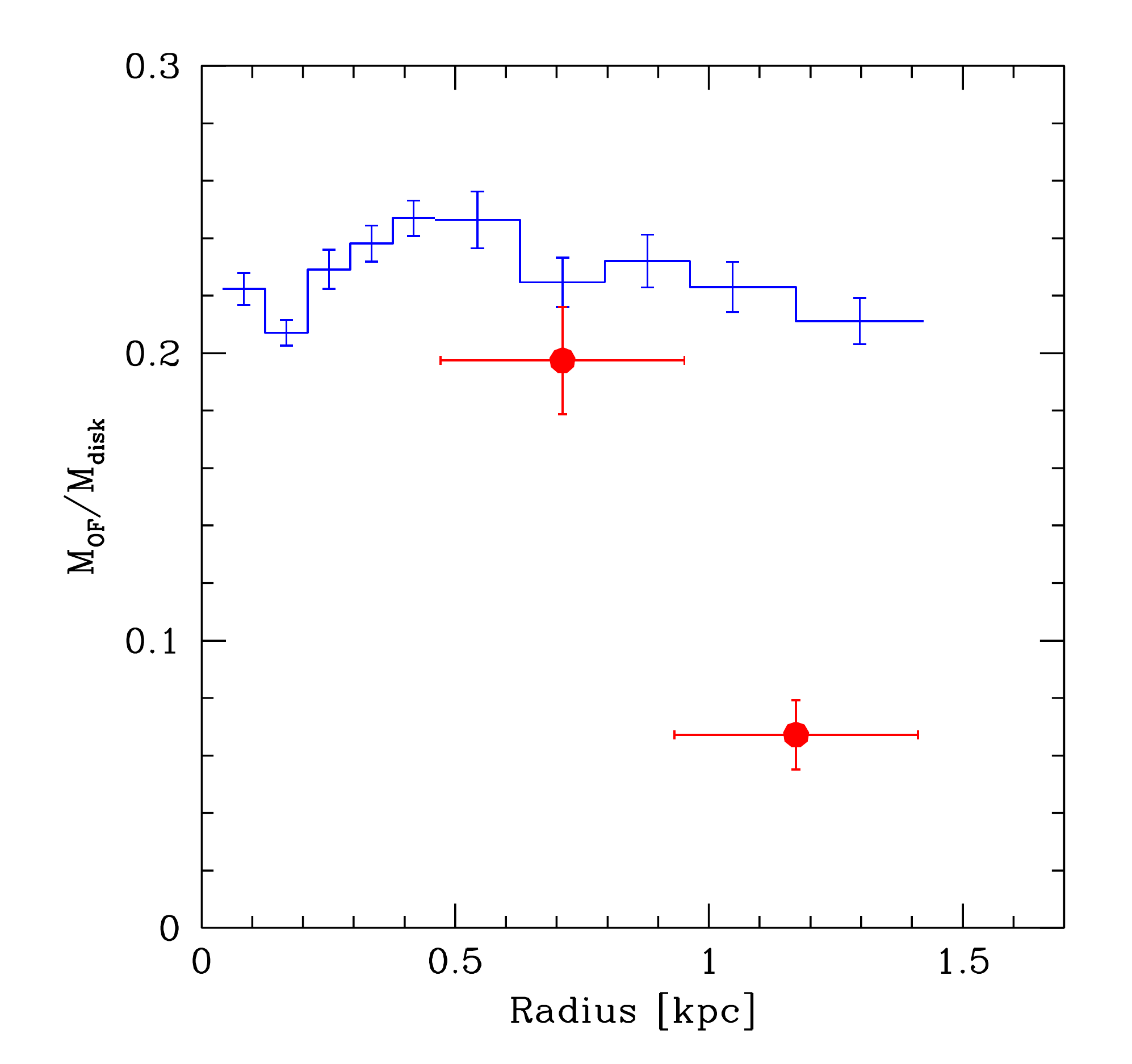}
\caption{Left panel: integral radial profiles (histogram) and profiles in annuli (points) of the outflow rate $\rm \dot M_{OF}$ (left axis), and $\rm v_{max}$ (right axis).
Middle panel: radial profile of $\rm \dot E_{kin,OF}$. Right panel:  radial profile of the outflow/molecular disk mass, $\rm M_{OF}/M_{disk}$. 
 }
 \label{mdotr}
\end{figure*}

Figure \ref{specs} compares the best fit CO(2-1) emission line
profiles integrated in a square region around the nucleus (using
DS1) with the integrated profiles from two adjacent square annuli at
increasing distances from the nucleus (using DS3). The rms is $\sim
1.3\times 10^{-3}$ Jy for all the spectra.  For the receding gas, we
detect high speed (800 km s$^{-1}$) gas out to $\simgt 1$ kpc, and a
deficit of gas with intermediate velocity (300-500 km s$^{-1}$) at
$\simgt$0.5 kpc.  For the approaching gas, this deficit is only seen
at $\geq$0.9 kpc.  The outflow mass rate $\rm \dot M_{OF}$, the $\rm
v_{max}$, the kinetic energy rate $\rm \dot E_{kin,OF}= 0.5\times \dot
M_{OF} \times v^2_{max}$, and the ratio of outflow mass and molecular
disk mass $\rm M_{OF}/M_{disk}$ are shown in Figure \ref{mdotr} as a
function of the distance from the nucleus (error-bars represent the
statistical errors only).  Specifically, the histograms represent
integral quantities out to a given radius, while the points represent
the local mass outflow rate in two annuli, computed by measuring the
mass density and the outflow mass within the annuli.  The integral
mass outflow rate $\rm \dot M_{OF}$ is $\approx 1000~ \rm
M_{\odot}~yr^{-1}$ within 400 pc from the nucleus, and 500-700 $\rm
M_{\odot}~yr^{-1}$ out to $\sim1$ kpc.  It is worth noting that the
local mass outflow rate is about 500 $\rm M_{\odot}~yr^{-1}$ within
$\sim800$ pc, while it drops to a few tens $\rm M_{\odot}~yr^{-1}$ at
$\simgt 1$ kpc.  The $\rm v_{max}$ and the integral $\rm \dot
E_{kin,OF}$ of the outflow remains nearly constant out to $\sim1$ kpc,
with $\rm \dot E_{kin,OF}=7-10\times 10^{43} erg \sim$ 1-2\% of the
AGN bolometric luminosity (Figure\ref{mdotr}, middle panel, see
Section 5 for a detailed discussion).  Finally, Figure \ref{mdotr},
right panel, shows that the outflow carries $\sim$0.2-0.25 of the
total disk mass out to $\sim1$ kpc, while the outflow mass drops to
less than 10\% of the disk mass at $\simgt 1$ kpc.

%\begin{figure*}
%\centering
%\includegraphics[scale=0.3]{massprofile.pdf}
%\caption{
% }
% \label{massprofile}
%\end{figure*}

\section{X-ray observations}

\subsection{X-ray data reduction}

During the last three years Mrk 231 has been target of new, sensitive
X-ray observations. Specifically, {\it Chandra} observed the galaxy
for 400 ks in August 2012 (Veilleux et al. 2014), while {\it NuSTAR}
has observed it twice, in August 2012 and May 2013 for a total of
about 70 ks (Teng et al. 2014,T14 hereafter).  These data have dramatically changed
our understanding of the X-ray emission from Mrk 231.  Previous
broadband, non-focusing X-ray observations performed with {\it BeppoSAX} 
and {\it Suzaku} detected a $\sim$3$\sigma$ excess in the
band above 10 keV which has been interpreted as evidence of nuclear
continuum emerging after transmission through a Compton thick absorber
(Braito et al. 2004), most likely with a variable covering factor
(Piconcelli et al. 2013).  This scenario has not been confirmed by the
unprecedented angular resolution ultra-hard ($>$10keV) X-ray {\it
  NuSTAR} observations presented by T14. They did not
report any hard X-ray excess and revealed that Mrk 231 is therefore
intrinsically X-ray weak, with a 2--10 keV luminosity of $4 \times
10^{42}$ erg s$^{-1}$. The best-fit model of the contemporaneous {\it
Chandra} $+$ {\it NuSTAR} spectrum consists of flat ($\Gamma \approx
1.4$) power-law continuum emission modified by a patchy, Compton-thin
absorber, plus a soft X-ray, starburst related, thermal emission.
Furthermore, the deep {\it Chandra} observation has revealed the
existence of a huge ($\sim$ 65 $\times$ 50 kpc) soft X-ray halo around
the central AGN which can be accounted for by two thermal emission
components with $k$$T$ $\sim$ 0.25 and 0.8 keV, respectively (Veilleux
et al. 2014).  Thanks to their high quality and sensitivity, these
datasets also allow a detailed search for highly ionized fast or
ultra-fast outflows seen in absorption against the nuclear X-ray
emission.

{\it Chandra} data were taken  from the CXC archive. Specifically, we combined
the 2012 long (400 ks, Observation ID 13947, 13948, 13949) observation with those performed
in 2000 and 2003 (153 ks in total, Observation ID 1031, 4028, 4029, 4030, Gallagher et al. 2005).
The combined data set has a total exposure time on source of 553 ks. 
Data were reduced using CIAO 4.5. We extracted a spectrum from a circular region of 3
pixel radius (1.5 arcsec) centered on the nucleus using the
tool {\sc dmextract}. 
A background spectrum was extracted from an
annulus with inner and outer radii 1 and 2 arcmin, respectively. In extracting the background, the regions of the front-illuminated detector have been masked. 
We verified that varying the background extraction regions, however, has little impact, because the background counts are a small fraction of the source counts 
in the spectral region of interest (a factor of $\sim1/500$). 
Response matrices were computed using the tools {\sc mkwarf} and {\sc mkrmf}.  
The spectral analysis was performed in the 0.5-10 keV energy range. 
Given the large number of counts and the requirement to use the $\chi^2$ statistics in our modeling, 
we binned the spectrum with a minimum of 40 counts/channel.
%Spectra were binned with a minimum of 40 counts/channel to
%allow the use of the $\chi^2$ statistics in the spectral fits. 

The {\it NuSTAR} data were reduced with the pipeline NuSTARDAS version
0.11.1 and CALDB version 20130509 with the standard settings (see T14 for details).  
The background counts are a factor of  $\sim1/10$ those of the source. 
Spectra were extracted for each observation and for the two {\it NuSTAR}
telescopes FPMA and FPMB, using a circular region of 1 arcmin radius.
Spectra were binned with a minimum of 40 counts/channel as for the {\it Chandra} data set. 
Spectral bins between 3 and 79 keV were used in the fits.
{\sc xspec} 12.8.0 was used for the analysis.

\subsection{Discovery of a nuclear Ultra-Fast Wind} 

\begin{figure}
\centering
\includegraphics[scale=0.37]{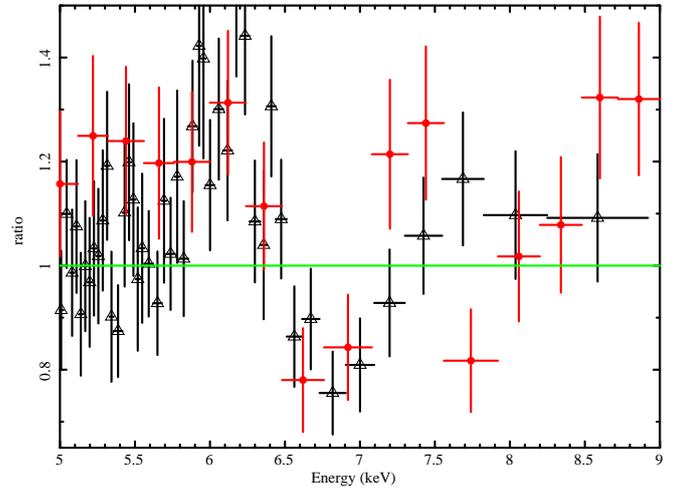}
\caption{The ratio between data the best model including two thermal
  components, a power law component and a narrow gaussian emission
  line at 6.18 keV (6.44 keV rest frame, iron K$\alpha$). Black triangles = {\it Chandra} data; red dots = {\it NuSTAR} data. 
A strong excess around 6 keV and a deficit of counts around 7 keV are seen. Note that the {\it NuSTAR} data, even taking into account the lower spectral resolution, follow the {\it Chandra} data.}
 \label{xraydata1}
\end{figure}

We exploited these data sets to constrain any nuclear wind.  Based on
T14 results, we first fitted the {\it Chandra} spectrum
and the four {\it NuSTAR} spectra with a model including Galactic
absorption along the line of sight, two thermal components, a power
law component and a narrow emission line component, both reduced at
low energies by photoelectric absorption (we used the {\sc xspec}
model {\sc zxipcf}, i.e. a model including a partial covering and
ionized absorber). The total $\chi^2$ of this fit is 498.9 for 468 deg
of freedom (dof). Figure \ref{xraydata1} shows the ratio between data
and model.  For plotting purposes the four {\it NuSTAR} spectra have
been added together.  Positive residuals are evident between 6 and 6.5
keV (6.2-6.7 keV rest frame, consistent with neutral Fe K$\alpha$
emission) while negative residuals are seen around 7 keV.  The feature
resembles a P-Cygni profile, similar to that recently found by Nardini
et al. (2015) in PDS456.  In the latter case the emission line peaks
however at $\sim7$ keV rest frame, suggesting highly ionized Fe
emission, unlike the case of Mrk~231.  The deficit of counts around 7
keV is also very well visible in Figures 4 and 5 of T14.  We replaced the narrow emission line with a
relativistic disk-line with inner and outer radii fixed at 10 and 1000
gravitational radii, and spectral index of the power emissivity law
fixed at -2.  The $\chi^2$ improves to 470.0 for 467 dof. The
equivalent width of the disk-line is 270 eV for the {\it Chandra}
spectrum and around 130-140 eV for the four {\it NuSTAR}
spectra. Residuals still show a deficit of counts around 7 keV.

\begin{figure*}
\centering
\includegraphics[scale=0.21]{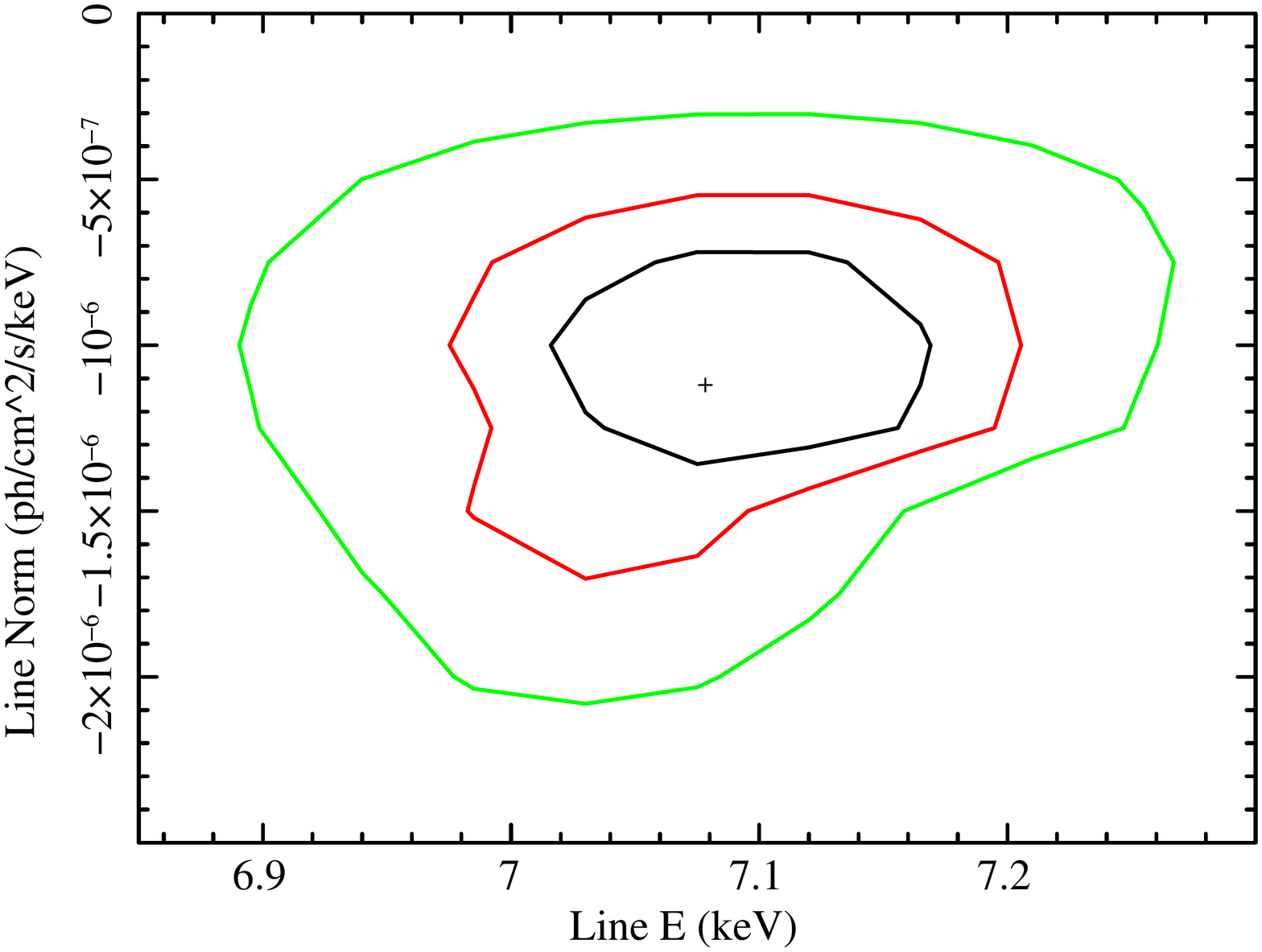}
\includegraphics[scale=0.21]{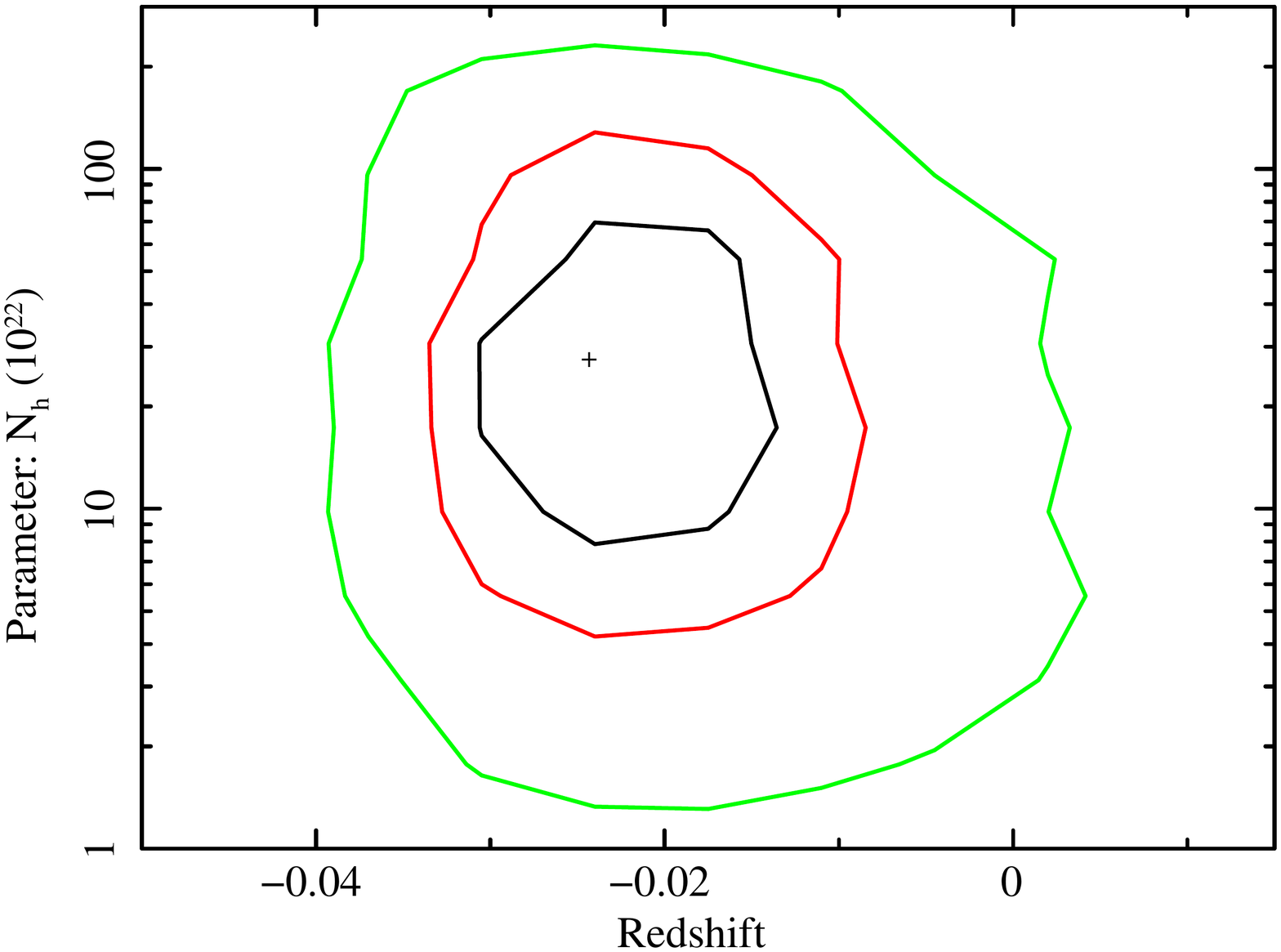}
\includegraphics[scale=0.21]{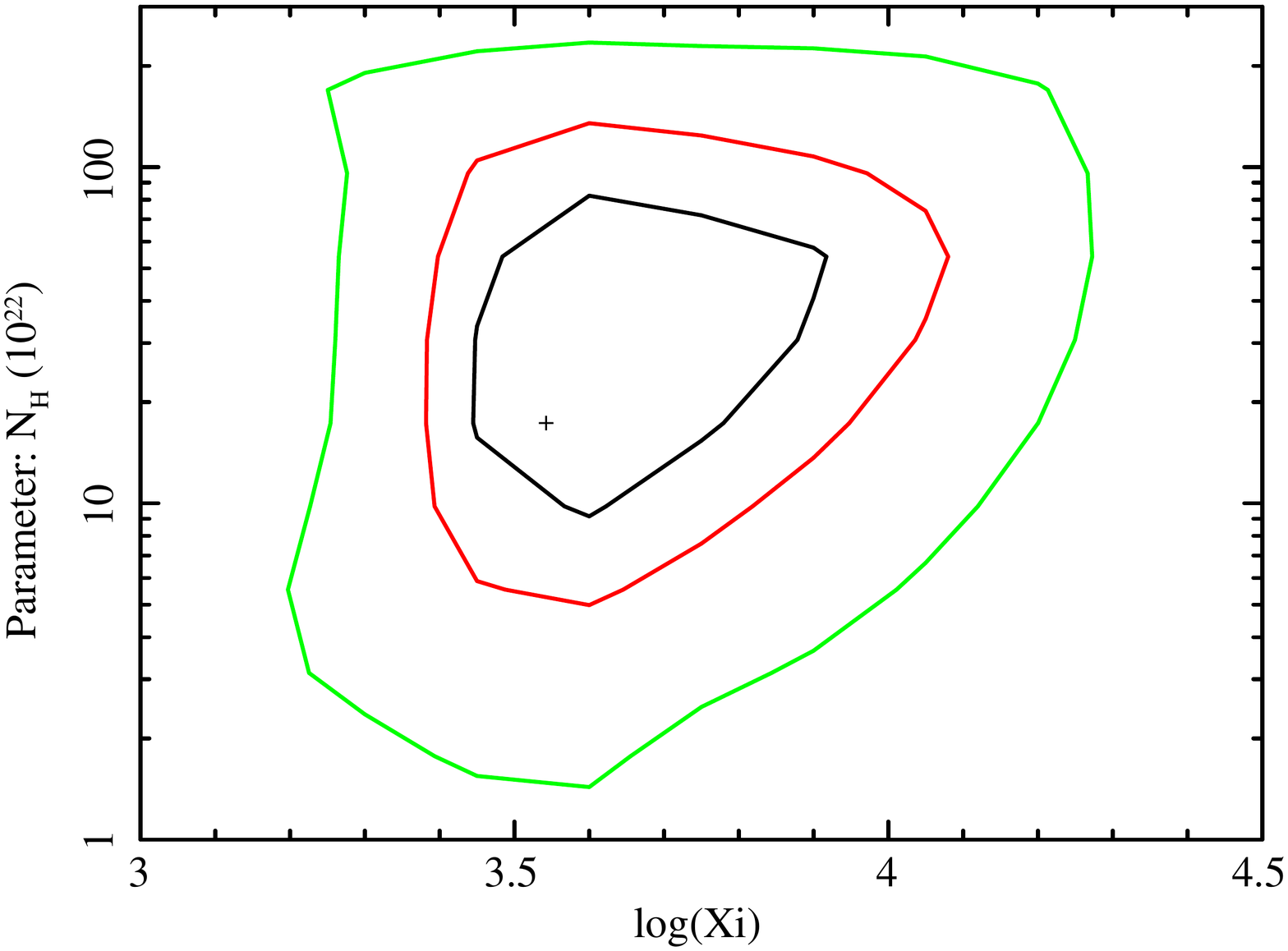}
\caption{Left panel: 67\%, 90\% and 99\% iso-$\chi^2$ confidence levels derived from the fitting of the merged X-ray spectra for the absorption
  line normalization and energy (left panel), for the column density, $\rm N_H$,
  and redshift of the highly ionized absorber (middle panel), and for $\rm N_H$ and
  ionization parameter, $\xi$, of the highly ionized absorber (right panel).}
 \label{xraydata2}
\end{figure*}

We then added to the model a Gaussian absorption line, to account for
this deficit of counts.  The $\chi^2$ further improves to 452.45 for
464 dof when the normalization of the gaussian absorption line is kept
linked in the five spectra. The $\Delta\chi^2$ between the model
without and with the gaussian absorption line is 17.55 for three
additional dof and no systematic residuals are seen in the iron line
complex.  Figure \ref{xraydata2} shows the $\chi^2$ contours for the
absorption line normalization and energy.  The detection of the
absorption line is significant at better than the 99.9\% confidence
level (or 3.5$\sigma$).  
To confirm the statistical significance of the absorption feature we run a Markov Chain Monte Carlo simulation by using the XSPEC \emph{chain} command. This generates a chain of set of parameter values describing the parameter probability distribution. 
The chain is converted into probability distribution using the XSPEC \emph{margin} command.

The $\Delta \chi^2$ between the models with and without the absorption line for the single {\it
  Chandra} spectrum is 11.6, while it is 7.3 between the two models
for the four NuStar spectra. The energy of the line is also similar in
the two instruments, $7.1\pm0.5$ keV for the {\it Chandra} spectrum
and $7.1^{+0.5}_{-0.4}$ keV for the {\it NuSTAR} spectra. We conclude
that the line is present with similar statistical significance in both
instruments. The equivalent width of the absorption line is about
twice in the NuSTAR spectra than in the {\it Chandra}
spectrum. Fitting simultaneously the {\it Chandra} and {\it NuSTAR}
spectra, but allowing the normalization of the absorption line to be
different in the {\it Chandra} and NuStar spectra, reduces the
$\chi^2$ to 448.24, i.e. the $\Delta \chi^2$ is only 4.2. For the sake
of simplicity, in the following analysis we keep the normalization of
the absorption line linked between the {\it Chandra} and {\it NuSTAR}
spectra.

To further test the robustness of the detection of the absorption line
at about 7.1 keV, we performed several additional spectral fits.
First, we limited the band used to fit the {\it NuSTAR} data to 30-40
keV, up to which the source is detected with a signal to noise better
than 3$\sigma$ in bins smaller than 10 keV. The results were perfectly
consistent with those obtained fitting the {\it NuSTAR} data in the
full band, up to 79 keV.  Second, we considered each single {\it
  Chandra} spectra (eight different observations) and fitted them
simultaneously. The results were consistent with those obtained
fitting the single {\it Chandra} spectrum obtained by adding together
the eight observations. Third, we fit the {\it Chandra} and {\it
  NuSTAR} data with a model including an absorption edge at energy
$\geq 7.1$ keV, rest frame.  The $\chi^2$ obtained linking the optical
depth $\tau$ of the edge between the {\it Chandra} and {\it NuSTAR}
data is 470.5 for 465 dof, and the resulting $\tau$ is consistent with
zero. Allowing the optical depth of the edge to be different in the
{\it Chandra} and {\it NuSTAR} spectra produces a $\chi^2$=458.3 for
464 dof, again much higher than the $\chi^2=448.24$ obtained using an
absorption line.

We fitted the {\it Chandra} and {\it NuSTAR} spectra with a model
  including two thermal components (T1 and T2), a power law nuclear
  component (PL) and a disk line component. The latter two components
  are modified by two ionized absorbers (a low ionization absorber to
  account for the cold-warm gas along the line of sight, and a highly
  ionized absorber to account for the fast wind). The $\chi^2$ is again
  good, 452.4 for 463 dof. The best fit parameters for all components
  are given in Table \ref{tab3}.
Figure \ref{xraydata2} shows the
$\chi^2$ contours for the highly ionized absorber N$_H$, ionization
parameter and redshift.
From the absorber redshift we derive a velocity of the absorbing gas
of $\rm v_{UFO}=-20000^{+2000}_{-3000}$ km s$^{-1}$. The
covering factor of the nuclear wind is consistent with being 1. 
 The statistics only allows to obtain a lower limit of 0.7 at
  the 67\% confidence level, as derived from the partial covering model (e.g. from the ratio of absorbed/unabsorbed power law, Table \ref{tab3}).
In principle, the wind covering factor may  be determined by a detailed analysis of the P-Cygni profile of the iron line feature, if the emission and the absorption come from the same expanding envelope (as in Nardini et al. 2015). Here we do not attempt such a detailed analysis, because the statistics does not allow us to distinguish between line emission from a wind and line emission from a relativistic accretion disk.   We limit ourselves to note that a prominent P-Cygni profile (e.g. strong emission in addition to strong absorption, Figure \ref{xraydata1}) points toward a large wind covering factor, provided that both features are produced by the same gas.

The comparison between our best fit model and the ÒpreferredÓ model of T14 is not straightforward because we used a simpler ionized/partial covering absorber model for the nuclear obscuration, while they adopted the ÒMYTORUSÓ model. Furthermore, we used a two temperature MEKAL model for the starburst component, while T14 use two temperature MEKAL model plus a power law component and a 6.7 keV Fe line to account for contribution of a population of nuclear High Mass X-ray Binaries (HMXB). 
Finally, we used a disk Fe K$\alpha$ line, while this component is automatically included in the MYTORUS model in T14.
The best fit spectral index and luminosity of the nuclear X-ray component are similar in the two fits, $\Gamma=1.47$ and $\rm L(2-10keV)=3\times10^{42}$ ergs/s  versus $\Gamma=1.4$ and $\rm L(2-10keV)\sim4\times10^{42}$ ergs/s.
Our parametrization of the starburst component provides a warm component (T=0.89 keV, similar to T14), plus a hot component (5.5 keV, which naturally accounts for the Fe 6.7 keV line). We tried to add a flat ($\Gamma=1.1$) power law component to account for a HMXB population, but this does not improve significantly the $\chi^2$, and its normalization is $<30$ times that of the AGN component (similarly to T14).
We fit a Compton thin absorber with column density $\sim4\times10^{22}$ cm$^{-2}$, very low ionization parameter, uniformly covering  the X-ray source, while T14 fit the data with two neutral absorber of similar total column densities.
In conclusion, despite significant differences in the details of the adopted models, the broad band X-ray spectral reconstruction is quite similar, and characterized by a) a main AGN power law component with a rather flat spectral index; b) a Compton thin, moderately ionized absorber of $\sim4\times10^{22}$ cm$^{-2}$, uniformly covering the source; c) a warm thermal component and a hot component (or a flat power law plus a 6.7 Fe line) to account for the nuclear starburst.
The detection of the absorption feature at about 7 keV is clearly not an artifact of the different spectral modeling, see discussion above, and it is well visible also in the residuals of Fig. 3 of T14.

  The other outflow parameters can be computed from the best fit
  parameters as follows.  The minimum distance of the outflowing gas
  can be estimated from the radius at which the observed velocity
  equals the escape velocity, $\rm r_{min}=2GM_{BH}/v_{UFO}^2 =
  5\times 10^{15}cm$ for a black hole mass of $8.7\times10^7$ M$_\odot$
  (Kawakatu et al. 2007) and the lowest value of the outflow velocity.
  The maximum distance $r_{max}$ of the outflowing gas can be computed
  from the definition of the ionization parameter $\rm \xi={L_{ion}
    \over n r^2} $,

\begin{equation}
r_{max}={L_{ion} \over N_H(min) \xi(min)}=3.0\times 10^{16} cm
\label{rmax}
\end{equation}

\noindent
where $L_{ion}\sim10^{43}$ ergs/s is the luminosity integrated in the 13.6 eV - 13.6 keV range.
The mass outflow rate at $r_{min}$ and $r_{max}$ can then be estimated using:

\begin{equation}
\dot M_{UFO}=0.8\pi m_p N^{UFO}_H v_{UFO} r f_g = 0.3- 2.1 ~M_\odot~ yr^{-1}
\label{mdotufo}
\end{equation}

\noindent
where $\rm N^{UFO}_{H}=2.7\times 10^{23}$ cm$^{-2}$, and $f_g$ is a
geometrical factor introduced by Krongold et al. (2007) and
statistically estimated $\sim2$ by Tombesi et al. (2010).  Should the
mass outflow rate be significantly larger than the value given above,
a much deeper absorption line should be visible in the X-ray spectra
(for reasonable values of $\xi$ and $v_{out}$). From the mass outflow
rate and velocity we can estimate the momentum flux, $\rm \dot
P_{UFO}=(0.4-2.7)\times10^{35}$ gr cm/s$^2$, corresponding to a $\rm
\dot E_{kin,UFO}=3.8\times 10^{43}-2.7\times 10^{44}$ erg~s$^{-1}$.  
This can be compared with the radiation momentum flux $\rm \dot P_{rad} =
L_{bol}/c$.  The bolometric luminosity estimated by Lonsdale et
al. (2003) and Farrah et al. (2003) is $5\times10^{45}$ erg~s$^{-1}$.
The momentum load is therefore $\rm \dot P_{UFO}/(\dot P_{rad}) =0.2-
1.6$.  Recently, Teng et al. (2014) measured an intrinsic X-ray
luminosity in the range 0.5-30 keV of $\sim 10^{43}$ erg~s$^{-1}$,
about 10 times smaller than previous estimates.  This would either
imply an unusually large X-ray bolometric correction ($\approx 1000$)
and hence a very large momentum load, or an overestimated AGN
bolometric luminosity based on mid-infrared data.

\begin{figure*}
\centering
\includegraphics[width=\textwidth]{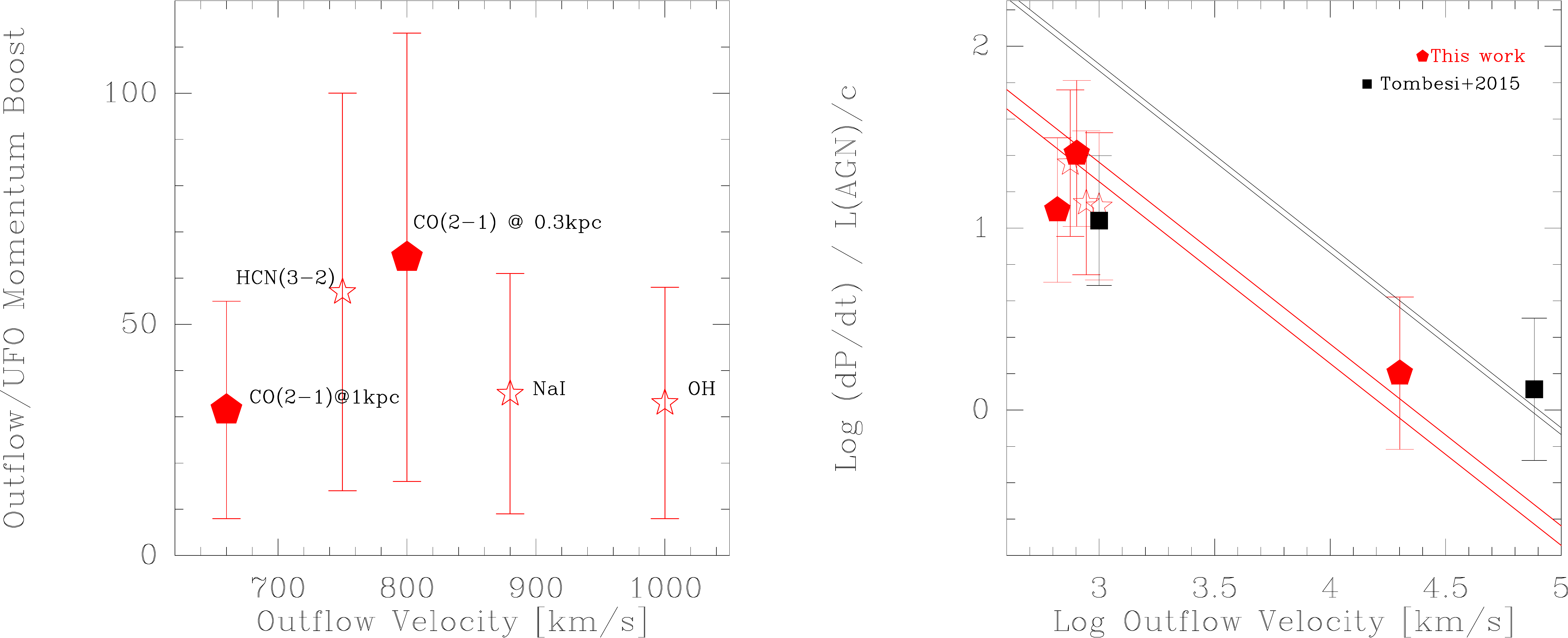}
\caption{Left panel: the {\it momentum boost} $\rm \dot P_{OF} / \dot P_{UFO}$ for the known molecular and atomic outflows of Mrk~231. 
References are Aalto et al. (2014) for HCN, Gonzalez-Alfonso et al. (2014) for OH, Rupke \& Veilleux (2011) for NaID, this work for CO.  
Right panel: the ratio of $\dot P_{OF}/ \dot P_{rad}$ versus velocity of the outflows of Mrk~231 (diamond: this work, stars: other works, see references above). 
Filled squares show the measurements for IRAS F11119+3257 (Tombesi et al. 2015). 
Red and black solid lines are the expectations for energy conserving outflows, $\rm \frac {\dot P_{OF}}{\dot P_{rad}} = \frac{v_{wind}}{v_{OF}}$, with $\rm v_{wind}=v_{UFO} \pm 1\sigma$ for Mrk~231 and IRAS F11119+3257, respectively.
}
\label{pload}
\end{figure*}

The detection of an UFO offers the unprecedented opportunity to
directly compute the momentum load of the large scale, spatially
resolved outflow with respect to that of the nuclear wind.  The $\rm
\dot P_{OF}$ of the CO(2-1) outflow depends on velocity and is in the
range $1.6-4\times 10^{36}$ erg cm$^{-2}$ for v=500 and 800 km
s$^{-1}$, respectively.  Figure \ref{pload}, left panel, shows the
momentum load, $\rm \dot P_{OF}/ \dot P_{UFO}$, versus the velocity of
the molecular and atomic large scale outflows detected in Mrk 231
(references for the data are in the caption of the figure).  The
values of $\rm \dot P_{OF}/ \dot P_{UFO}$ range from $\approx20$ to
$\approx100$, i.e.  strongly suggesting that the outflow does not
conserve the momentum, {\it if} the UFO is identified with the inner
semi-relativistic wind pushing the outer molecular outflow (Lapi et
al. 2005, Zubovas \& King 2012, Faucher-Giguere \& Quataert 2012).
Figure \ref{pload}, right panel, shows the $\dot P_{outflow}/ \dot
P_{rad}$ versus velocity of the outflows of Mrk~231 (red symbols).
Black symbols are the results of Tombesi et al. (2015) for IRAS
F11119+3257, a galaxy exhibiting both an X-ray UFO and an OH outflow,
as revealed by Herschel spectroscopy (Veilleux et al. 2013).  The
solid lines represent the expectations for energy-conserving outflows
with nuclear wind velocity $v_{UFO}\pm 1\sigma$ and with covering
factor $f=1$ for both the nuclear winds and the molecular outflows.

\begin{table}
\caption{Best fit model to the {\it Chandra} and {\it NuSTAR} merged X-ray spectra}
\label{tab3}
\begin{center}
\begin{tabular}{lcc}
\hline
Parameter & Best fit value & Units \\
\hline
KT(1)   & 0.89$\pm0.05$                  & keV \\
Normalization(1) & (4.9$\pm0.5)\times10^{-6}$  & $\rm ph~ cm^{-2}~s^{-1}~ keV^{-1}$ \\
KT(2)   & 5.5$\pm0.7$                 & keV \\
Normalization(2) & ($5.7\pm2.1)\times10^{-5}$  &  $\rm ph~ cm^{-2}~s^{-1}~ keV^{-1}$ \\
N$_H$(1)       & (4.2$\pm0.3) \times10^{22}$    & $\rm cm^{-2}$  \\
Covering Factor(1)  & 0.97$^{+0.03}_{-0.04}$                &  \\
log$\xi$(1)    & -0.54$^{+0.16}_{-0.04}$                 &  $\rm erg~cm~s^{-1}$ \\
N$_H$(2)       & (2.7$^{+3.6}_{-1.5}) \times10^{23}$    & $\rm cm^{-2}$  \\
Covering Factor(2)  & 1.0$_{-0.3}$                &  \\
log$\xi$(2) & 3.55$^{+0.2}_{-0.1}$       &  $\rm erg~ cm~s^{-1}$ \\
$\Gamma$(PL)    & 1.47$\pm0.1$               &  \\
Normalization(PL)    & (1.9$\pm0.2) \times10^{-4}$  &  $\rm ph~ cm^{-2}~s^{-1}~ keV^{-1}$ \\
Disk Line Energy    & 6.05$\pm0.04$                & keV \\
Disk Inclination       & 35$\pm5$                & deg \\
Normalization(Line)  & (3.0$\pm0.5)\times10^{-6}$   &   $\rm ph~ cm^{-2}~s^{-1}~ keV^{-1}$ \\
\hline
$\chi^2$ (dof) & 452.4 (463)& \\
\hline
\end{tabular}
\end{center}
\end{table}

\begin{figure*}
\centering
\includegraphics[width=10cm, angle=0]{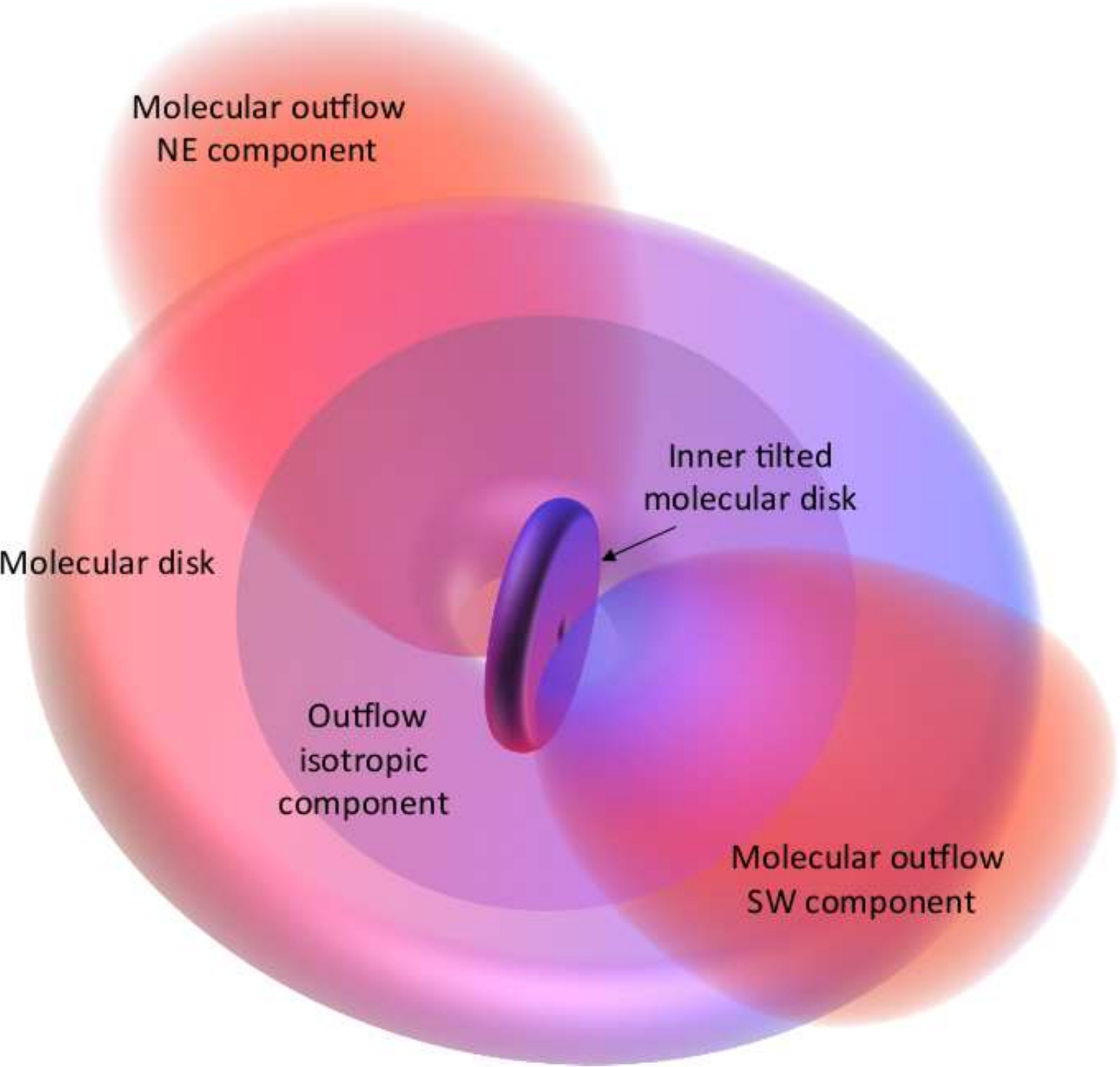}
\caption{3D cartoon of the molecular disk and outflow geometry in Mrk~231. The line of sight to the observer is perpendicular to the page. 
Red and blue colours represent  the receding and approaching material for each component, respectively. 
The molecular disk consists of an inner disk, tilted by 58 deg on the line of sight, with a scale radius of 200 pc,  plus and outer component with inclination 36 deg and scale radius 400 pc, with the line of nodes oriented in the east-west direction approximately.
The molecular outflow is rendered by a wide-angle biconical-like component plus an isotropic one, represented here as a sphere for displaying purposes.  For a detailed description of the cartoon see Section 5.
}
\label{sketch}
\end{figure*}

\section{Discussion and conclusions}

%\begin{enumerate}[i]

 {\it (i) Molecular disk}. 
We present the best spatial resolution and sensitivity CO map of the molecular disk of Mrk~231 obtained so far. 
In addition to the previously known main rotation of the disk, which occurs along nearly an east-west direction, we find evidence for a tilted inner nuclear disk on scales of 100 pc,  more inclined on the line of sight. 
The CO gas distribution can be modeled by an outer exponential ring with radius 400 pc, $\rm PA= -12$ deg, and inclination $i=36\pm10$ deg, plus an 
inner exponential disk with outer radius 200 pc, $\rm PA= 84$ deg, and $i=58$ deg on the line of sight. 
The inner, warped component matches the size, PA and inclination of the vibrationally excited HCN(3-2) v=1,f (Aalto et al. 2014). 
%the the high excitation rotating ('quiescent') component of the OH absorption observed by Herschel (Gonzalez-Alfonso et al. 2014)
The parameters of the inner tilted disk are also consistent with the OH mega-maser emission seen with the VLBI, which traces a disk with radius 30 to 200 pc, PA 56 deg and inclination 56 deg (Klockner et al. 2003). 
We can hardly estimate the mass of the inner tilted CO disk, but this detection shows that in addition to very dense gas traced by HCN(3-2) $v_2=1f$, gas with density of the order $\rm 10^5~cm^{-3}$ is also present in the very inner regions. 
The existence of a warped inner disk with radius 170-250 pc was suggested also by Davies et al. (2004) based on stellar absorption studies. 
They pointed out that the warp should first occur in the gaseous phase and so be transferred to the stars formed in situ.

\smallskip

{\it (ii)  Molecular outflow.}
The CO(2-1) high angular resolution data indicate that the molecular outflow has a size of at least 1 kpc. The bulk of both receding and approaching outflowing gas are located within $\sim$400 pc from the nucleus, and peak $\sim 0.2$ arcsec south-west off the nucleus. 
Extended, redshifted emission with lower surface brightness is seen north-east off the nucleus out to $\sim1$ kpc (see Fig. 3). 

The outflow is seen in all examined directions around the AGN (see Fig. 6). 
It is more prominent, however, along the south-west to north-east direction, suggesting a wide-angle likely biconical geometry. 
%approximately perpendicular to the inner molecular disk. 
The approaching gas shows a gradient in projected velocity, with larger negative speed at the position of the AGN and decreasing velocity south-west off the nucleus. %consistent with a decelerating outflow, 
%as proposed by some models (Roth et al 2012, Faucher-Giguere \& Quataert 2012) and as required to explain the OH complexes (Gonzalez-Alfonso et al. 2014). 
This allows to exclude a geometry where the approaching outflow expands with constant speed within an uniformly filled cone with axis located in the plane of the sky (as in this case no speed gradient would be observed) nor with axis parallel to our line of sight (in this case the maximum projected speed would be centered on the nucleus and decrease outwards symmetrically). 
An outflow expanding with uniform velocity between the observer and the molecular disc, in a cone with axis  inclined with respect to our line of sight, would instead produce a velocity gradient as the observed one for the approaching gas, but it would also imply a similar gradient for the receding part, which instead is not observed. 
The projected velocities can be probably qualitatively be explained by an outflow with conical or lobe-like geometry with a slightly larger receding component (based on Fig. 2 and 3). However a firm and consistent  picture explaining both velocity fields seen in Fig. 9 cannot be simply ascribed to orientation effects and probably inhomogeneities and asymmetries in the outflow can account for them. 

%This, together with the velocity gradient seen for the blue-shifted component, may indicate that the molecular outflow is decelerating with increasing distance from the AGN, and that we are seeing the approaching portion of a cone located in front of the molecular disk. 

We also suggest that the redshifted emission seen north-east off the nucleus is tracing the receding part of the cone, which is located behind the molecular disk with respect to the observer. 
Interestingly, Krabbe et al. (1997) detected at approximately the same location a narrow Pa$\alpha$ emission approaching with a speed of 1400 km s$^{-1}$ with respect to the systemic one. 
The molecular outflow centroids, traced by the red and blueshifted wings of CO(2-1), are consistent with those seen in  HCN (Aalto et al. 2014). 
%The south-west to north-east orientation of the diffuse outflow component appears inconsistent with the east-west orientation seen by CARMA in CO(1-0) (Alatalo et al. 2015). 
%While we cannot exclude that this inconsistency is due to the lower sensitivity and angular resolution of the CARMA data, it is also possible that CARMA data maps a more diffuse and extended component than that in this work. 

We find that $\rm \dot M_{OF}$ decreases with increasing distance from the nucleus in all directions (see Fig. 11), while the radially integrated $\rm v_{max}$ stays roughly constant out to $\simgt1$ kpc.  
This implies, based on equation \ref{fluid}, that either the outflow average density, $\rm \rho_{OF}$, or the outflow filling factor decrease from the nucleus outwards approximately as $r^{-2}$ (or slightly steeper). 
This suggests different possible scenarios: (i) a large part of the gas with velocity of 300-600 km s$^{-1}$ leaves the flow during its expansion.
If the molecular clouds are pressure-confined, they will dissolve out in the wind, and CO may be efficiently photo dissociated by the UV radiation, since self shielding will be strongly reduced at low densities. Atomic carbon, either in neutral or ionized phase, can
then, in principle, continue expanding out to larger distances at a constant velocity, as it is observed in SDSSJ1148+5251, where most of
the fast [CII] is located far from the AGN, and extended on scales of a few tens kpc (Cicone et al. 2015).  
Intriguingly, broad wings extending up to $\pm 1000$ km/s can be seen in both [CII] 158 $\mu$m and [NII] 205 $\mu$m emission lines observed by {\it Herschel} in Mrk 231 (Fischer et al. 2010). Unfortunately, Herschel PACS  does not provide sufficient angular resolution to assess whether the fast [CII] and [NII] gas are more extended than CO. A promising alternative would be to use the forbidden 3P fine structure line of atomic carbon ([CI] line at rest frame frequency 492.161 GHz), which is accessible by ALMA, on southern analogs of Mrk 231.
 (ii) The opening angle subtended by the paths of least resistance is relatively narrow, or (iii) only the highest velocity gas has been able to reach distances $\simgt 1$ kpc. 
The latter would imply an age of the outflow of $\sim1$ Myr (assuming a constant expansion), which is 1-10\% of a typical AGN duty cycle.
The rate of kinetic energy carried by the outflow is $\sim$ 1-2\% of the AGN bolometric luminosity, and stays approximately constant out to 1 kpc. The decrease of the local $\rm \dot E_{kin,OF}$ at larger radii can be explained as a consequence of decreasing $\rm \dot M_{OF}$ and constant $\rm v_{max}$. 

A 3D rendering of the geometry of the molecular disk and outflow is shown in Figure \ref{sketch}. 
The molecular disk consists of an inner tilted disk plus and outer component with approximately an east-west orientation. 
The observed direction of the rotation for these two rotating components is shown by the colour scale (red for the receding, blue for the approaching material). 
The molecular outflow is sketched out by a wide-angle biconical-like component plus a spherical one. 
The two cone-like components have been placed in order to match the south-west to north-east elongation seen in the CO(2-1) maps.
The southern cone is located between the observer and the disk, inclined with respect to the line of sight, and shows an approaching and a brighter receding gas emission.
The northern cone is located behind the molecular disc, and it is mainly receding, as testified by the redshifted velocities observed at this position. 
Since we cannot draw firm conclusions on the expansion velocity along the plane of the sky, 
we mention the possibility that the isotropic outflowing component can also be explained by receding and approaching outflows along the line of sight.
It is worth noting that the neutral gas outflow reported by Rupke \& Veilleux (2011) extends over a much larger region
($\sim$3 kpc) than that of the molecular gas traced by CO.

\smallskip

{\it (iii) Molecular outflow and disk global energetics.} 
A complementary way to assess whether the outflow has a significant impact on the molecular disk is comparing their energetics, keeping in mind, however, that these energy estimates are affected by large uncertainties.  
The total energy of the  molecular disk (quiescent component) is given by the sum of the gravitational potential energy, the rotational energy, the turbulent energy and the thermal one.
 The gravitational potential energy is $\rm E_{grav}=G M_{dyn} M_{gas}/R$, 
 where $\rm M_{dyn}$ and $\rm M_{gas}$ are the dynamical and gas masses, 
 G is the gravitational constant and R is the radius of the molecular disk (400 pc from our fit). 
 The gas mass within this radius is $\rm M_{gas}=1.8\times 10^9~M_{\odot}$.
The dynamical mass, modulus the inclination of the disk, is $\rm M_{dyn}$$ sin^2(i)=5.5\times10^8~\rm M_{\odot}$, which gives $\rm E_{grav}~$$sin^2(i)=1.85\times 10^{56}$ erg.  

The rotational energy of the molecular disk is $E_{rot} = \frac{1}{2} M_{gas} v^2=2.1 \times10^{57}$ erg, where $v=345$ km s$^{-1}$ is the rotational velocity of the disk. 
The contribution of the turbulent energy is small and can be expressed as: $E_{turb} = \frac{1}{2} M_{gas}$d$v^2=2.9 \times10^{55}$ erg, where d$v$ is the turbulent velocity dispersion in the disk (which is likely of the order of $\sim40$ km s$^{-1}$). 
The thermal energy can be expressed as $\rm E_{therm}= n~$$k$$\rm~T~V= 1.2\times10^{52}$ erg, assuming a temperature of the gas of $\rm T=60~ K$ based on CO(1-0) observations, an average gas density $\rm n=3600~ cm^{-3}$, and a disk with radius 400 pc and thickness 23 pc (Downes \& Solomon 1998).
The total energy budget of the molecular disk is therefore $\rm E_{disk}=2.6-8.3\times10^{57}$ erg,  adopting a disk inclination of 36 and 10 deg, respectively. 

The kinetic energy of the outflow is given by its bulk motion.  
Based on the mass of the CO(2-1) outflow, we find $\rm E_{kin,OF}=2.35\times10^{57}$ erg, adopting an average velocity of 800 km s$^{-1}$ and an outflow mass of $\rm 3.7\times10^{8}~M_{\odot}$
The turbulent contribution is negligible if we assume $dv_{turb,OF}=dv_{turb,disk}\sim 40$ km/s. 
On the other hand if  in the outflowing molecular clouds $dv_{turb}\sim250$ km/s (e.g. Williamson et al. 2014), the turbulent energy can contribute an additional $\sim10\%$ to the outflow energy budget. 
Given all the uncertainties in these estimates, we conclude that the total energy of the outflow  $\rm E_{OF}$ is of the same order as $\rm E_{disk}$.

%The kinetic energy of the outflow is therefore 2-4 times smaller than the total energy content of the molecular disk. 
%seems, therefore, obsolete. 
%Our findings show that the molecular outflow can perturb and displace large gas mass within a distance of up to $\sim 1$ kpc away from the nucleus.
%According to our outflow maps, the outflow seems to escape {\it preferentially} in a least resistance path, i.e. nearly perpendicular to the inner molecular disk plane with a wide opening angle.  This supports the prediction of Faucher-Giguere \& Quataert (2012). 
%Even in the case where the energy of the outflow is released isotropically, the outflow would intercept a small fraction of the thin molecular disk. 
%Accordingly, the hypothesis that the massive molecular outflow is able to disrupt the entire molecular disk and expel most of the gas from the galaxy is not supported by our result, unless the energy of the outflow is released mainly in the molecular disk plane, which however seems excluded by our data. 
%  but suggest that it can hardly disrupt the molecular disk and/or eject most of the gas reservoir outside the host galaxy. 

\smallskip

{\it (iv) UFO and molecular outflow energy and momentum.}
We find a ratio $\rm \dot E_{kin,UFO}/\dot E_{kin,outflow}=0.4-3.8$, suggesting that most of the nuclear wind kinetic energy is transferred to mechanical energy of the kpc scale outflow, which is thus undergoing an energy conserving expansion, as predicted by the most popular theoretical  models (Faucher-Giguere \& Quataert 2012, Zubovas \& King 2012). 
In particular, the prediction of models for energy-conserving outflows is $\rm \dot E_{kin,UFO}=$$f$$\rm \times \dot E_{kin,OF}$, where $f$ is the fraction the nuclear wind energy that is deposited to the large scale outflow, and ranges from 0.5 to 1 in the models of Faucher-Giguere \& Quataert (2012) and  Zubovas \& King (2012), respectively.    
We stress that our work does not make any assumption on $f$. In fact, we measure 

\begin{displaymath}
\rm \frac {\dot P_{OF}}{\dot P_{UFO}} \approx  \frac{v_{UFO}}{v_{OF}}
\end{displaymath}

from which we derive $f\sim1$ (see also Figure \ref{pload}). 
The mass loading factor, defined as $f_l=\sqrt(\dot M_{OF}/\dot M_{UFO})$ in Zubovas \& King (2012), is $\approx 20$ at the outer boundary of the molecular outflow (1 kpc), which also supports the two stage acceleration scenario.  
We find that $\rm \dot E_{kin, UFO}/L_{bol,AGN} = 1-5\%$ and $\rm \dot E_{kin,OF}/L_{bol,AGN} = 1-3\%$, in agreement with the requirements typically assumed by the most popular feedback models (Lapi et al. 2005, Menci et al. 2008, Di Matteo 2005, Gaspari et al. 2011, 2012). 

The  momentum load of the nuclear wind, $\rm \dot P_{UFO}/(L_{bol,AGN}/c)=0.2-1.6$, agrees with the predictions for radiatively accelerated winds with scattering optical depth $\sim 1$ (King \& Pounds 2003).   
These observations offered the first opportunity to compare the momentum boost of a {\it spatially resolved} outflow with respect to the nuclear wind. 
Specifically, we find $\rm \dot P_{OF}/\dot P_{UFO}=[16-113]$  at 1 kpc based on CO(2-1), and in the range $8-100$ for the outflows traced by HCN, OH and NaID, although with large uncertainties (Figure \ref{pload}). 
Accordingly, this result supports the two phase outflow acceleration mechanism scenario, whereby the momentum of the large scale outflow is boosted compared to that of the nuclear semi-relativistic wind. 
We remark that the nucleus of Mrk~231 is radiating close to the Eddington limit ($\rm L_{bol,AGN}/L_{Edd} = 0.46$ for a black hole mass of $8.7\times 10^{7}~\rm M_{\odot}$), matching again the requirements of models for driving massive outflows. 

Tombesi et al. (2015) recently found a similar result for the galaxy IRAS F11119+3257, which hosts both an UFO and a massive molecular outflow detected through the OH absorption line by Herschel (Veilleux et al. 2013). They derived  $\rm \dot P_{OF}/\dot P_{rad}\approx 12$, estimating $f=0.2$. 
%For IRAS F11119+3257, the spatial extension of the molecular outflow is inferred through indirect arguments. 
It is worth noting that both Tombesi et al. (2015) and this work strengthen each other, supporting the AGN-driven energy-conserving outflows scenario and providing constraints to the models of AGN feedback. 

Sensitive and high angular resolution observations with ALMA and NOEMA are needed to further constrain the physics of quasar-driven outflows, and their impact in galaxy transformation. 
%Further mapping of the Mrk~231 molecular outflow will be possible with the NOEMA enhanced 8/10 antenna array in a few years from now. 
In particular, constraining any large-scale molecular outflow in quasars with well-studied UFO (e.g. PDS~456, PG~1211+143, Nardini et al. 2015, Pounds 2014) is a compelling experiment to be pursued with ALMA. 
On the other hand, understanding the impact of outflows on the cosmological evolution of galaxies  
requires  a systematic approach, i.e. a search for nuclear molecular outflows out to $z\sim2$ in unbiased, mass-selected samples.

\begin{acknowledgements}
CF thanks the IRAM staff for support in observations and data analysis, particularly Dennis Downes, Roberto Neri, Sergio Martin-Ruiz, Jan Martin Winters, Melanie Krips, and Sabine Koenig. 
We thank Susanne Aalto, Santiago Garcia-Burillo, and Andrea Ferrara for stimulating discussions. 
CF gratefully acknowledges financial support from PRIN MIUR 2010-2011, project ``The Chemical and Dynamical Evolution of the Milky Way and Local Group Galaxies'', prot. 2010LY5N2T. 
CF and FF acknowledge financial support from INAF under the contract PRIN-INAF-2011.  
EP and AB acknowledge financial support from INAF under the contract PRIN-INAF-2012.  
SV acknowledges financial support from NASA grant G02-13129X.
Data used in this work will be made available from http://cdsweb.u-strasbg.fr.
\end{acknowledgements}

\end{document}